\documentclass[manuscript,nonacm]{acmart}
\usepackage{xurl}
\usepackage{listings}
\lstdefinestyle{prompt}{basicstyle=\footnotesize\ttfamily,breaklines=true,
  breakatwhitespace=true,breakindent=0pt,postbreak={},prebreak=\mbox{$\hookleftarrow$},columns=fullflexible,
  keepspaces=true,frame=single,framesep=2mm,xleftmargin=2mm,xrightmargin=2mm,
  showstringspaces=false,upquote=true}
\usepackage{amsmath}
\usepackage{booktabs}
\usepackage{graphicx}
\usepackage{array}
\usepackage{threeparttable}
\usepackage{longtable}
\usepackage{seqsplit}
\usepackage{pgfplots}
\pgfplotsset{compat=1.18}
\usetikzlibrary{patterns}
\usepackage{algorithm}
\usepackage{algpseudocode}
\usepackage{float}
\usepackage{placeins}

\renewcommand\footnotetextcopyrightpermission[1]{}
\acmConference[Under review]{Under review}{}{}

\title{What Language Models Add to Healthcare-Equity Rankings, and What They Could Have Known}

\author{Haotong Fu}
\email{mfu@racap.com}
\author{Xiyue Zhang}
\email{vzhang@racap.com}
\author{Jack Rose}
\email{jk.kr19842@gmail.com}
\author{Ding Sun}
\email{dsun@racap.com}
\author{Erich Scheller}
\email{escheller@racap.com}
\author{Eric Rose}
\email{erose@racap.com}
\affiliation{%
  \institution{RA Capital Management}
  \city{Boston}
  \country{USA}
}
\makeatletter
\let\acm@mkauthors@i@orig\@mkauthors@i
\def\@mkauthors@i{\let\MakeUppercase\@firstofone\acm@mkauthors@i@orig}
\makeatother

\makeatletter\renewcommand{\@titlefont}{\sffamily\bfseries\fontsize{14}{17}\selectfont}\makeatother
\makeatletter\newcommand\eatperiod{\@ifnextchar.{\@gobble}{}}\makeatother
\begin{document}
% ---- AI4F body (assembled into paper_icaif.tex) ----

\begin{abstract}
In healthcare equities, much of what moves a single name arrives as text: filings, trial and regulatory records, news and analyst commentary. We evaluate how screened non-market features and language-model representations improve market-based equity rankings for US healthcare firms over 2020--2025. Adding balance-sheet and attention features improves the resulting hybrid's price- and liquidity-neutral rank information coefficient (IC) at one, three and twelve months. At twelve months, neutral IC rises from 0.063 to 0.093, a 48\% increase, while raw IC rises from 0.289 to 0.298. Language models contribute most clearly at three months: among company-months with scientific-text coverage, a current embedding model raises the hybrid's neutral IC from 0.046 to 0.056. The three-month increment is significant at conventional unadjusted levels ($t$ 2.8; paired 95\% bootstrap interval $[+0.006, +0.015]$) and remains exploratory. A larger post hoc gain occurs in months containing trial or regulatory records, consistent with text complementing prices and financial statements at that horizon. Across all text, pre-2020 and current representations yield small positive three-month increments in both neutral and raw IC but no established twelve-month gain. Much of their ranking performance survives within-company time shuffling, consistent with a persistent company component. Separate rater-cutoff comparisons find negligible pre-cutoff advantages for current models over the pre-2020 rater at one and three months that do not shrink after the cutoffs; twelve-month hindsight attribution remains inconclusive. At twelve months, balance-sheet signals lead after rallies and in volatile markets, while attention leads after declines. Retrospective state-dependent blending improves on a static blend without an established advantage over the full hybrid. Results apply to a survivor-only sector sample and remain unadjusted for the full exploratory search.
\end{abstract}

\maketitle
\pagestyle{plain}

\section{Introduction}
\label{sec:intro}

Language models now sit inside equity-ranking pipelines as raters of news, summarizers of filings and writers of the explanations an analyst reads. Reported gains from such components are common \citep{lopezlira2023chatgpt, chen2023expected}, and so is the worry that a model trained on text through 2025 scores a 2021 article knowing how the story ended \citep{glasserman2023lookahead, sarkar2024lookahead, ludwig2025llm}. We combine component comparisons with matched-text cutoff tests within one ranking system, using matched rows within each comparison.

This paper evaluates one such system on both counts. The system ranks US healthcare and life-science equities monthly over 2020--2025 from a contest-selected random forest on fourteen price-volume features (the \emph{market score}), blended by a fold-specific multiple-testing screen and a ridge head with 187 non-market columns: balance sheets, scientific and filing events, analyst coverage and ratings, and news and social text scored by FinBERT and by a two-stage language-model chain (GPT-5.4-mini summaries rated by Claude Sonnet 4). A companion descriptive layer writes a short narrative per company and horizon under a rulebook that forbids forecasting. The ranking's economics are reported elsewhere; here every component is placed on one ledger, model provenance is documented, and selected rating and forecasting components are tested for cutoff-related changes.

Four questions organize the evaluation, in the order the abstract states them. Eight hypotheses, H1 to H8, were specified before their reported runs, with exploratory extensions identified separately; they are listed with their dates and outcomes in Table~\ref{tab:prespec} (chronology in Table~\ref{tab:chrono}). \textbf{(1) What does the ranking learn from, and what does a language model add?} A component ledger (Table~\ref{tab:ledger}) measures each channel's paired gain over the market-only ranking, and five rankings are then compared on the same rows (Table~\ref{tab:models}, Figure~\ref{fig:models}): market-only, market plus a language-model text ranking, market plus screened non-market signals (the hybrid), the hybrid plus the text ranking, and text alone. \textbf{(2) Where does the text help, and why?} The text ranking is built from every non-numerical input the system holds, read by a pre-2020 model and by a current one under the same downstream protocol, and decomposed by source, horizon and event window against two placebos. \textbf{(3) Is it hindsight?} The pre-2020 rater and the frontier raters score identical text before and after the frontier models' training cutoffs, with a placebo pair of two cutoff-clean raters (training data predating the 2020--2025 ranking sample; the 2018--2019 rating-panel overlap is disclosed in Section~\ref{sec:leak}); a prompted frontier forecaster runs blind and revealed; the narrative layer is audited mechanically alongside the LLM judge that audits it. \textbf{(4) When does each source work?} A leave-one-year-out regime blend tests whether balance-sheet and attention signals alternate by market state; its hypothesis, H5, formalizes a pattern first seen in September 2026.

The contribution is a component-level evaluation within one sector, one benchmark and six years: numerical and attention features improve the ranking; scientific text provides an exploratory three-month lead, while an additional twelve-month text gain is not established; matched-text comparisons show negligible pre-cutoff frontier-minus-GPT-2 gaps at one and three months that do not shrink after the cutoffs, and do not identify hindsight bias at twelve; and the relative contributions of balance-sheet and attention signals vary with market conditions in retrospective analyses. Section~\ref{sec:system} describes the system and the ledger, Section~\ref{sec:clean} the rankings compared and where text helps, Section~\ref{sec:leak} the hindsight tests, Section~\ref{sec:state} the market states, Section~\ref{sec:related} related work and Section~\ref{sec:conclusion} the limits, the conclusion and the recommended model.

\section{The System and Its Component Ledger}
\label{sec:system}

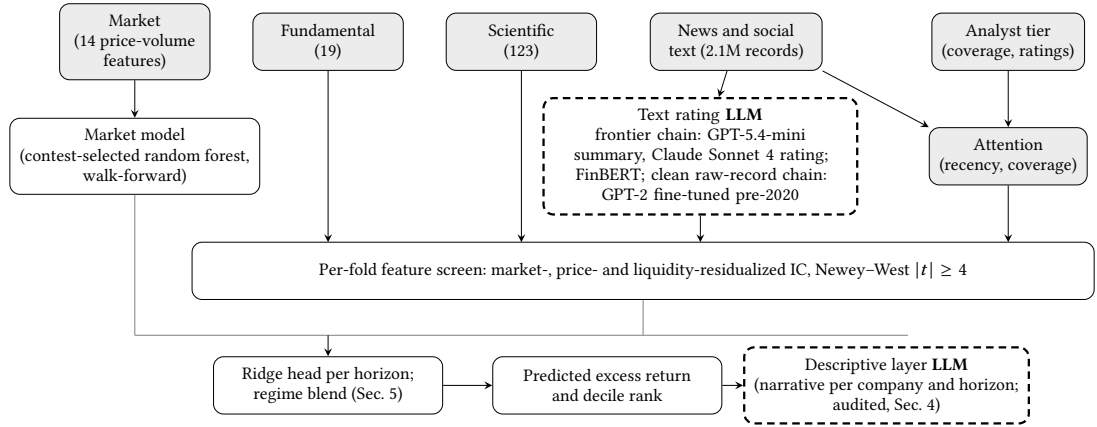
\begin{figure}[!ht]
\centering
\Description{Flowchart of the system: structured data boxes feed a market model and a per-fold feature screen, then a ridge head and regime blend produce the ranking; language-model stages are tagged LLM and drawn with dashed borders.}
\resizebox{0.95\textwidth}{!}{%
\begin{tikzpicture}[box/.style={draw, rounded corners, align=center, font=\footnotesize, inner sep=4pt, minimum height=0.8cm},
  llm/.style={box, fill=white, densely dashed, line width=0.9pt}, ai/.style={box, fill=white}, raw/.style={box, fill=gray!15}, >=stealth]
\node[raw, minimum width=2.1cm] (market) at (0,8)    {Market\\(14 price-volume\\features)};
\node[raw, minimum width=2.1cm] (fund)   at (2.7,8)  {Fundamental\\(19)};
\node[raw, minimum width=2.1cm] (sci)    at (5.4,8)  {Scientific\\(123)};
\node[raw, minimum width=2.4cm] (sent)   at (8.4,8)  {News and social\\text (2.1M records)};
\node[raw, minimum width=2.1cm] (an)     at (12.2,8) {Analyst tier\\(coverage, ratings)};
\node[ai, minimum width=3.2cm] (mkmodel) at (0,6.4) {Market model\\(contest-selected random forest,\\walk-forward)};
\node[llm, minimum width=3.8cm, text width=4.1cm] (sentq) at (7.9,6.4) {Text rating \textbf{LLM}\\frontier chain: GPT-5.4-mini\\summary, Claude Sonnet 4 rating;\\FinBERT; clean raw-record chain:\\GPT-2 fine-tuned pre-2020};
\node[raw, minimum width=2.1cm] (att) at (12.2,6.4) {Attention\\(recency, coverage)};
\draw[->] (market) -- (mkmodel); \draw[->] (sent) -- (sentq); \draw[->] (an) -- (att); \draw[->] (sent.south east) -- (att.north west);
\node[ai, minimum width=12.6cm] (screen) at (7.1,4.8) {Per-fold feature screen: market-, price- and liquidity-residualized IC, Newey--West $|t|\geq4$};
\draw[->] (fund.south) -- (2.7,5.2); \draw[->] (sci.south) -- (5.4,5.2); \draw[->] (sentq.south) -- (7.9,5.2); \draw[->] (att.south) -- (12.2,5.2);
\draw[gray] (mkmodel.south) -- (0,3.9); \draw[gray] (screen.south) -- (7.1,3.9); \draw[gray] (0,3.9) -- (10.8,3.9);
\node[ai, minimum width=3.2cm] (meta1) at (2.7,3.2)  {Ridge head per horizon;\\regime blend (Sec.~\ref{sec:state})};
\node[ai, minimum width=3.2cm] (meta3) at (6.6,3.2)  {Predicted excess return\\and decile rank};
\node[llm, minimum width=3.4cm] (meta12) at (10.5,3.2) {Descriptive layer \textbf{LLM}\\(narrative per company and horizon;\\audited, Sec.~\ref{sec:leak})};
\draw[->] (2.7,3.9) -- (meta1.north); \draw[->] (meta1) -- (meta3); \draw[->] (meta3) -- (meta12);
\end{tikzpicture}}
\caption{The system with its component ledger. Boxes tagged \textbf{LLM} are stages run by a language model. Dashed border: language-model stages (text rating in two chains, the narrative layer). Solid border, white: non-language machine learning (market block, screen, heads, regime blend). Grey: structured input data. The ranking is the output of the white boxes; the dashed boxes enter it only through screened columns, and no frontier-chain polarity column is ever admitted; the extracted-facts channel is admitted in two folds, inertly.}
\label{fig:system}
\vspace{-11pt}
\end{figure}

\paragraph{Universe, data and horizons.} A company enters if its primary four-digit FactSet NAICS code is one of thirteen healthcare and life-science classes; 4{,}868 companies, January 2010 to June 2026, with each month's cross-section reduced to names priced at or above \$1 in the top half by trailing dollar volume (28.5\% of the universe). Targets are log excess returns over XBI at 21, 63 and 252 sessions. Five source pipelines feed the system (Figure~\ref{fig:system}; volumes and admissions in Table~\ref{tab:volumes}): market (10.1M daily price-volume records), sentiment (2.1M records from eight feeds in three tiers: news, social, analyst), fundamental (33k XBRL filings), scientific (25k events) and macroeconomic. Every non-market family is joined on its public-availability date and carried forward; the sentiment, fundamental and scientific feeds were collected in June 2026 against the companies listed then, so every non-market gain below is a within-survivor estimate (Section~\ref{sec:conclusion}).

\paragraph{Market block, screen and head.} The market block is a random forest chosen by open contest among sixteen architectures on 2015--2018 walk-forward windows (contest results not reproduced here); for test year $Y$ it is trained on 2010 through $Y-2$ and the resulting market score is the \emph{market-only} ranking. At each horizon a screen tests every non-market column on the fold's training rows: the column is rank-residualized within month against the market score and then against four price and liquidity controls (log price, dollar volume, volatility, Amihud illiquidity), and admitted if its monthly IC clears a Newey--West $|t|\geq4$ with 5\% coverage, fifteen names a month and 24 months of signal. The gate was fixed in the September 2026 analysis of the same system, before any row here was run, and sits slightly above the within-fold Bonferroni value of 3.92 for 561 tests; the closest frontier polarity miss, $|t|$ 3.96, would clear at 3.92 in one fold--horizon pair (coverage gates permitting), and its gain was not computed. A ridge head ($\alpha=10$, inputs demeaned within month) blends the admitted columns with the market score; companies without enough admitted coverage keep their market-score rank exactly, and the covered companies are re-ordered among themselves within the set of rank positions they collectively held under the market score, so a covered company can cross deciles while the uncovered companies' ranks and the month's distribution of predicted values are unchanged. Nine of 187 columns are admitted in at least one fold--horizon pair, none of them a text score.

\paragraph{Two metrics.} Every gain is a paired month-by-month difference of two rankings' Spearman rank ICs on held-out 2020--2025 months, with Newey--West $t$ (lags $=$ horizon) and, for twelve-month claims on 65 overlapping months, a circular block bootstrap (block twelve, 10{,}000 draws). The \emph{raw} IC is the unadjusted ranking association; the \emph{price- and liquidity-neutral} IC is computed within each month $m$ as follows: with $r_i$ the percentile rank of the ranking and $\mathbf{c}_i$ the percentile ranks of the four controls (log price, dollar volume, volatility, Amihud illiquidity), the residual $e_i = r_i - \hat\alpha - \hat{\boldsymbol\beta}^{\prime}\mathbf{c}_i$ is taken from a least-squares fit on that month's cross-section, and the neutral IC is the Spearman correlation $\rho_S(e, y)$ of the residuals with the realized excess returns $y$, which are not residualized. The metric is therefore the Spearman correlation between control-residualized score ranks and realized returns; because Spearman re-ranks the residuals it is not a standard partial correlation, and ``neutral IC'' is this paper's operational label for that control-adjusted rank IC. It measures what a ranking knows beyond price and liquidity level. The admission screen is stricter than the metric: it residualizes a candidate column on the market score as well as on the four controls before testing its IC. The neutral metric is the research metric; the raw metric is reported beside it. Several results are computed on subsamples (text-covered rows, scientific-text rows, the 300-company forecasting sample, the 2018--2026 rating panel), each with its own baseline IC; Table~\ref{tab:panels} lists them so that no increment is read against another sample's baseline.

\paragraph{The ledger.} Table~\ref{tab:ledger} re-runs the screen and head with the candidate pool restricted to one channel at a time, under the identical per-fold rule, so each row is that channel's own hybrid model against the same market-only base. The sentiment pipeline's 45 columns are split by what they measure, not by their source: \emph{attention}, twenty counts of record recency and coverage that contain no text content; \emph{polarity}, twenty machine-scored text columns (FinBERT per source and tier, the Claude and FinBERT ratings of the GPT-5.4-mini summaries); \emph{analyst ratings}, five recommendation scores; and \emph{extracted facts}, nine new columns built for this paper from the event type, materiality and themes GPT-5.4-mini extracts under its summary prompt. Rows are paired gains against one base and do not add. The full model's candidate pool is the original 187 columns; the twelve GPT-2-rated columns and the nine fact columns enter only their own rows.

\begin{table}[!htbp]
\centering
\caption{Component ledger: paired gain of each channel's own hybrid model over the market-only ranking, 2020--2025, by horizon; price- and liquidity-neutral rank IC (research metric) with Newey--West $t$ (lags $=$ horizon) and the raw rank IC gain. Base row: the market-only IC itself. ``0'' marks channels in which no column clears the screen in any of the six folds at that horizon, so the gain is zero by construction; for these the post hoc magnitude bound is the largest $|$IC$|$ of any of the channel's individual columns, after the screen's two residualizations on the evaluation rows, that a 95\% interval cannot rule out (columns passing the 24-month, 5\% coverage gates); it bounds column magnitudes, not the gain of every possible text model, and no equivalence margin was specified in advance. 72, 72 and 65 months. Panel A gives the neutral gains and bounds, Panel B the raw gains.}
\label{tab:ledger}
{\footnotesize\setlength{\tabcolsep}{4pt}
\textbf{A. Price- and liquidity-neutral rank IC gain ($t$) and post hoc magnitude bound}\par\smallskip
\begin{tabular}{@{}llrrrl@{}}
\toprule
Component & Type & 1m & 3m & 12m & Bound (1m / 3m / 12m) \\
\midrule
Market-only ranking (base IC) & model & 0.020 (3.5) & 0.027 (3.9) & 0.063 (8.3) & \\
\midrule
$+$ Fundamental (19 balance-sheet columns) & data & 0 & $+0.009$ (1.8) & $+0.022$ (2.8) & \\
$+$ Scientific events (123) & data & 0 & 0 & $-0.001$ ($-1.6$) & \\
$+$ Attention: recency and coverage (20) & data & $+0.008$ (2.8) & $+0.013$ (3.9) & $+0.021$ (3.3) & \\
$+$ Analyst recommendation scores (5) & data & 0 & 0 & 0 & 0.033 / 0.047 / 0.082 \\
$+$ Text polarity, frontier chain (20) & LLM & 0 & 0 & 0 & 0.042 / 0.059 / 0.065 \\
$+$ GPT-2-rated columns: 11 scores, 1 recency (12) & LLM & 0 & 0 & 0 & 0.044 / 0.081 / 0.063 \\
$+$ Extracted facts: event type, materiality, themes (9) & LLM & 0 & 0 & $+0.000$ (0.6) & 0.050 / 0.083 / 0.115 \\
\midrule
Full model (all families) & model & $+0.008$ (2.8) & $+0.017$ (3.6) & $+0.030$ (3.9) & \\
\bottomrule
\end{tabular}
\par\medskip
\textbf{B. Raw rank IC gain ($t$)}\par\smallskip
\begin{tabular}{@{}llrrr@{}}
\toprule
Component & Type & 1m & 3m & 12m \\
\midrule
Market-only ranking (base IC) & model & 0.132 (7.4) & 0.195 (6.6) & 0.289 (6.9) \\
\midrule
$+$ Fundamental (19) & data & 0 & $-0.002$ ($-1.3$) & $+0.007$ (1.8) \\
$+$ Scientific events (123) & data & 0 & 0 & $-0.000$ ($-1.4$) \\
$+$ Attention: recency and coverage (20) & data & $-0.000$ ($-0.1$) & $+0.001$ (0.9) & $+0.005$ (3.9) \\
$+$ Analyst recommendation scores (5) & data & 0 & 0 & 0 \\
$+$ Text polarity, frontier chain (20) & LLM & 0 & 0 & 0 \\
$+$ GPT-2-rated columns (12) & LLM & 0 & 0 & 0 \\
$+$ Extracted facts (9) & LLM & 0 & 0 & $+0.000$ (1.6) \\
\midrule
Full model (all families) & model & $-0.000$ ($-0.1$) & $-0.001$ ($-0.6$) & $+0.009$ (2.7) \\
\bottomrule
\end{tabular}}
\par\medskip\begin{minipage}{\linewidth}\footnotesize\raggedright\setlength{\parskip}{1pt}
\par Months with a positive neutral gain for the full model: 50 of 72 at one month, 37 of 48 at three (37 of 72 over all months; the 24 months of two folds in which nothing is admitted are excluded from the first count) and 54 of 65 at twelve. Block-bootstrap 95\% intervals for the three positive twelve-month rows: fundamental $[+0.007, +0.039]$, attention $[+0.010, +0.034]$, full $[+0.014, +0.046]$. Admitted columns by fold and horizon are in Table~\ref{tab:volumes}; the analyst-ratings and polarity rows are zero in all eighteen fold--horizon pairs (closest misses $|t|$ 3.67 and 3.96 on training rows), and the facts row admits one column in each of two twelve-month folds (financing-themed in 2021, regulatory-themed in 2025) and moves the ranking by $+0.0001$, its bounds wide because the summaries are dense only from 2025. The bounds are post hoc, on evaluation rows, and are upper bounds on column magnitude, not estimates of any model's gain. Ablation without the recency column: the full model's neutral gain is 0, $+0.009$ and $+0.022$ at one, three and twelve months (raw IC 0.132, 0.193 and 0.296 against the market-only 0.132, 0.195 and 0.289); at one month the ablated ranking is the market-only ranking, since no non-market column remains admitted.
\end{minipage}
\end{table}

The full model raises the market-only ranking's neutral IC by about half at twelve months (0.063 to 0.093) and by a third to two thirds at the shorter horizons; on the raw metric only the twelve-month gain survives (0.289 to 0.298, a rise of three percent). Price trends come first: the market-only ranking's neutral IC, 0.063 at twelve months, exceeds every gain. Balance-sheet numbers and attention follow, each adding about $+0.02$ neutral IC at twelve months; attention contributes at all three horizons, fundamentals most clearly at twelve ($+0.009$, $t$ 1.8, at three). Content adds no measurable gain at either model generation under this screen: no analyst-rating or polarity column, frontier or GPT-2-rated, is admitted, the extracted facts enter two twelve-month folds without a measurable gain, and post hoc bounds on their individual residualized columns place their twelve-month neutral IC below 0.063--0.115 (Table~\ref{tab:ledger}, Panel A, last column); the bounds do not exclude every practically meaningful effect, and they bound columns, not models. The predictive attention channel needs no language model to compute: it counts records and analysts. That is itself an evaluation result about where the language models in Figure~\ref{fig:system} earn their place. The ledger sees text only as scores and counts. The next section reads the text itself, with a pre-2020 model and a current one, and asks what a language-model ranking adds to each of the rankings above.

\section{What the Language Model Adds, and Where}
\label{sec:clean}

The ledger's text rows are scores: a rater reads each record and returns a number, and the screen tests the number. This section instead reads the text itself. Every non-numerical input the system holds is turned into one vector per company-month by a frozen language model, a walk-forward head turns the vectors into a ranking, and that ranking is compared with, and added to, the market-only ranking and the hybrid on the same rows. The comparison is run twice, with a pre-2020 model and with a current one, so that the result does not rest on a model that could have seen the sample's outcomes.

\paragraph{Rankings on the same rows.} Every non-numerical input the system holds is used and nothing else: for each company-month, the text of all records dated in the trailing 30 days, most recent first, tagged by source, up to 6{,}000 characters, from news, social posts and scientific text (sources in Appendix~\ref{app:data}); 522{,}968 records make 74{,}575 company-months for 1{,}136 companies from 2015, and text exists for 41.5--43.7\% of evaluation rows. Balance-sheet numbers have no text and are excluded by design; no frontier-written summary enters. Two frozen models turn each company-month's text into one vector under the same downstream training protocol, with model-specific truncation, pooling and normalization (Appendix~\ref{app:tables}): GPT-2 small (2019, 768 dimensions, a causal language model) and Qwen3-Embedding-0.6B \citep[2025, 1{,}024 dimensions, an embedding model;][]{zhang2025qwen3embedding}. They differ in size, objective, training data and retained input (512 tokens against 6{,}000 characters) as well as in date, so the comparison is between two representations, not a controlled test of model generation. For each test year a ridge head is fitted on rows through $Y-2$ to the within-month demeaned excess return and applied to year $Y$, and the predictions are ranked within month; nothing is prompted. Two shuffling controls test the source of predictive association: shuffling vectors across companies within each month breaks the text--company link, and shuffling each company's vectors across months keeps the company and breaks the timing of its news. Each text ranking is added to the market-only ranking and to the hybrid as a within-month rank blend, with a plain average or with a weight selected on the other five evaluation years, on both metrics; because those years include years after the held-out one, the weighted blends are retrospective cross-validated comparisons, not a walk-forward test, though the selected weight is constant across folds for the all-text hybrid blends at three and twelve months (0.1) and for the scientific-text hybrid blend at three months (0.2); the market-plus-text weights vary. Pre-specified (H6): the text ranking's neutral IC is indistinguishable from zero at every horizon and it adds nothing to the numeric model.

\begin{figure}[!b]
\centering
\begin{tikzpicture}
\begin{axis}[ybar, bar width=5.5pt, width=0.92\linewidth, height=5.2cm, ymin=0, ymax=0.125,
  ylabel={Neutral rank IC, text rows}, ylabel style={font=\footnotesize}, tick label style={font=\footnotesize}, scaled y ticks=false, yticklabel style={/pgf/number format/fixed, /pgf/number format/precision=2}, ytick={0,0.025,0.05,0.075,0.1,0.125},
  symbolic x coords={1m,3m,12m}, xtick=data, enlarge x limits=0.22,
  legend style={font=\scriptsize, at={(0.02,0.98)}, anchor=north west, legend columns=2, draw=none, fill=none},
  ymajorgrids, grid style={gray!30}, axis line style={gray!60}]
\addplot[fill=gray!15, draw=black] coordinates {(1m,0.019) (3m,0.022) (12m,0.057)};
\addplot[fill=gray!55, draw=black] coordinates {(1m,0.031) (3m,0.043) (12m,0.110)};
\addplot[fill=black, draw=black] coordinates {(1m,0.030) (3m,0.047) (12m,0.113)};
\addplot[fill=white, draw=black, postaction={pattern=north east lines}] coordinates {(1m,0.031) (3m,0.046) (12m,0.112)};
\addplot[fill=white, draw=black, densely dashed] coordinates {(1m,0.005) (3m,0.030) (12m,0.039)};
\addplot[fill=white, draw=black, densely dotted] coordinates {(1m,0.000) (3m,0.019) (12m,0.034)};
\legend{Market-only, Hybrid (market $+$ non-market), Hybrid $+$ text (pre-2020 model), Hybrid $+$ text (current model), Text alone (pre-2020), Text alone (current)}
\end{axis}
\end{tikzpicture}
\caption{Six of the rankings of Table~\ref{tab:models} by horizon (the market-plus-text rows are omitted), price- and liquidity-neutral rank IC on the evaluation rows with text. Adding either language-model text ranking to the hybrid moves it by $+0.003$ or less at twelve months; the text rankings alone reach a third of the hybrid's IC at twelve months, and the identity placebo (Table~\ref{tab:texttower}) matches them.}
\Description{Grouped bar chart of neutral rank IC at one, three and twelve months for six rankings; the hybrid and the hybrid-plus-text bars are indistinguishable, the market-only bar is about half the hybrid at twelve months, and the text-alone bars are about a third.}
\label{fig:models}
\end{figure}
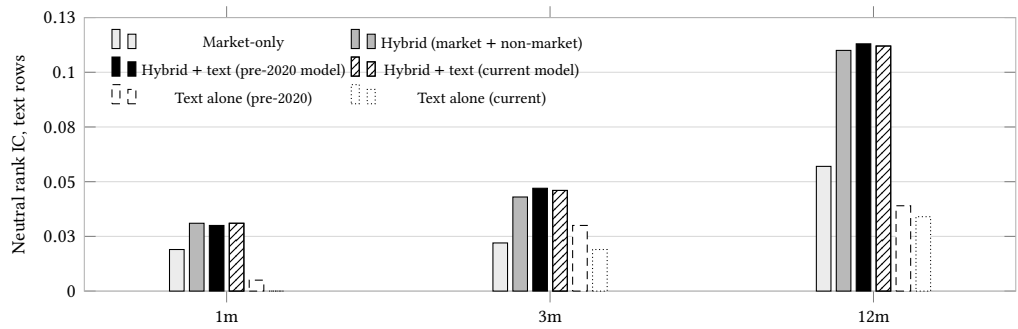

Table~\ref{tab:models} and Figure~\ref{fig:models} give the rankings side by side on the rows that have text (the full detail, with placebos and both blend forms, is Table~\ref{tab:texttower} in Appendix~\ref{app:tables}), and the answer is the same for both representations. The text rankings predict: at twelve months the current model's neutral IC is $+0.034$ ($t$ 3.4) and the pre-2020 model's $+0.039$ ($t$ 4.1), while shuffling the text across companies eliminates the positive association (the pre-2020 model's cross-company placebo is $-0.010$, $t$ $-2.6$). But the within-company shuffle (the identity placebo), which keeps every company's text and destroys its timing, predicts about as well, and better in five of the six cells (Table~\ref{tab:texttower}; the exception is the pre-2020 model at twelve months, $+0.036$ against $+0.039$). This suggests that much of the skill is persistent company-level language, the name, sector, stage and size vocabulary a language model encodes together with whatever recurs in a company's coverage. The shuffle does not separate identity from every persistent textual feature, and no identity-only baseline was run, but it shows that the month's content adds little beyond them. That persistent component is largely already in the numbers (Table~\ref{tab:models}): adding either text ranking to the hybrid at a learned weight of 0.1 changes its twelve-month neutral IC by $+0.003$ (paired intervals $[-0.002, +0.008]$ and $[-0.001, +0.007]$), while a plain average lowers it by 0.035--0.040. The one exception is the pre-2020 model at three months, $+0.004$ neutral ($t$ 2.0, $[+0.002, +0.005]$) and $+0.002$ raw ($t$ 3.1). H6's first half is not met, since both twelve-month ICs are distinguishable from zero; its second half, which named no horizon, is not met at three months ($+0.004$ and, for scientific text, $+0.010$, intervals excluding zero) and not rejected at twelve, where neither representation establishes an additional gain over a ranking that already knows the company from its prices and filings. Per-fold fine-tuning of the encoders was not run.

\begin{table}[!b]
\centering
\caption{Rankings on the same rows: evaluation rows with text (41.5--43.7\% of all rows; 72, 72 and 65 months), 2020--2025. Price- and liquidity-neutral rank IC by horizon, raw rank IC at twelve months, and for the blends the paired change against the corresponding base ranking, market-only or hybrid (Newey--West $t$). Text rankings are added to the market-only ranking and to the hybrid at the weight chosen leave-one-year-out on the other five years (0.1 at twelve months in every fold for the hybrid; 0.1--0.3 for the market-only ranking). The scientific-text block is computed on the rows with scientific text (33\%), whose baselines differ.}
\label{tab:models}
{\footnotesize
\resizebox{\linewidth}{!}{\begin{tabular}{@{}lrrrrl@{}}
\toprule
Ranking (text rows) & 1m & 3m & 12m & 12m raw & $\Delta$ vs base, neutral 1m / 3m / 12m ($t$) \\
\midrule
Market-only & 0.019 & 0.022 & 0.057 & 0.244 & \\
Market-only $+$ text ranking, pre-2020 model & 0.020 & 0.029 & 0.058 & 0.241 & vs market-only: $+0.002$ (1.2) / $+0.007$ (1.0) / $+0.001$ (0.2) \\
Market-only $+$ text ranking, current model & 0.018 & 0.022 & 0.060 & 0.245 & vs market-only: $-0.001$ ($-1.6$) / $0.000$ (0.0) / $+0.003$ (0.7) \\
Hybrid: market $+$ screened non-market signals & 0.031 & 0.043 & 0.110 & 0.265 & \\
Hybrid $+$ text ranking, pre-2020 model (GPT-2) & 0.030 & 0.047 & 0.113 & 0.266 & vs hybrid: $-0.002$ ($-1.7$) / $+0.004$ (2.0) / $+0.003$ (1.5) \\
Hybrid $+$ text ranking, current model (Qwen3-Embedding) & 0.031 & 0.046 & 0.112 & 0.266 & vs hybrid: $0$ (weight 0) / $+0.003$ (1.5) / $+0.003$ (1.0) \\
Text alone, pre-2020 model & 0.005 & 0.030 & 0.039 & 0.120 & \\
Text alone, current model & 0.000 & 0.019 & 0.034 & 0.113 & \\
\midrule
\multicolumn{6}{@{}l}{\emph{Rows with scientific text (33\%)}} \\
Hybrid & 0.032 & 0.046 & 0.105 & 0.269 & \\
Hybrid $+$ scientific-text ranking (current model) & 0.032 & 0.056 & 0.108 & 0.270 & vs hybrid: $0$ (weight 0) / $+0.010$ (2.8) / $+0.003$ (1.2) \\
Scientific text alone & 0.003 & 0.040 & 0.035 & 0.125 & \\
\bottomrule
\end{tabular}}}
\end{table}

\paragraph{Where the text helps, and why.} Six transformations of the text ranking, specified before running (Appendix~\ref{app:prespec}), each change the lens: a change vector (this month's news), a residual, a magnitude target, tail counts, coverage gaps, and splits by source and by attention; Table~\ref{tab:lenses} (Appendix~\ref{app:lenses}) reports them for the current model, and the pre-2020 model gives the same verdict on each lens for which it was run, all but the news-alone and social-alone arms (its values are in the table's note). A lens shows content only if its neutral IC clears $|t|\geq3$ and is not matched by the identity placebo.

Five of the six show no content, and the sixth shows one lead (Table~\ref{tab:lenses}). The change vector, the direct test of ``this month's news'', predicts nothing once a company's usual text is removed; the residual head, the magnitude head and the attention split recover the persistent company signal under another name, with the identity placebo matching or exceeding the residual and magnitude heads for the current model (for the pre-2020 model the residual head exceeds its placebo yet adds only $+0.002$ to the hybrid; the attention split has no placebo); the tail columns reverse sign between training and evaluation rows, as a coverage count does; text does not fill coverage gaps. The lead is scientific text at three months. Re-embedding the panel one source at a time, the ranking built from scientific text alone (33\% of evaluation rows) has neutral IC $+0.040$ ($t$ 5.4), clearing the standalone threshold the lens rule set in advance, though its identity placebo nearly matches it ($+0.035$; the margin carries no interval). Added to the hybrid at a leave-one-year-out weight of 0.2 in every fold, it raises the hybrid's three-month neutral IC by $+0.010$ ($t$ 2.8, bootstrap $[+0.006, +0.015]$) and its raw IC by $+0.003$ ($t$ 2.2), where the identity placebo's blend adds $-0.001$. The increment is significant at conventional unadjusted levels and remains exploratory; the increment does not appear at twelve months or one, and the contrast motivates testing whether time-specific scientific information explains it rather than establishing that reading. A post hoc subgroup comparison (L6c) shows a larger gain in the 16\% of company-months whose window holds a trial-registry, FDA or advisory-committee record, where a separately fitted blend selects scientific text alone and improves neutral IC by $+0.064$ over the subgroup's hybrid baseline of $-0.023$ ($t$ 2.9; the identity placebo's blend adds $+0.024$), against $+0.005$ elsewhere; it is a subgroup model comparison, not a decomposition of the aggregate increment, and its raw increment is small. The horizon pattern, a gain at three months and none at one or twelve, is consistent with content about trials and regulatory actions that resolve over a quarter, an interpretation on about 49 names a month rather than an established explanation. The system already held scientific information as 123 structured columns, which the screen admitted only occasionally with no reliable aggregate increment; the same records, read as text and tested under a different rule on a third of the rows, show a lead the screened counts did not, and the matched comparison on the same rows is the test for later data. News and social text alone add nothing beyond the persistent signal in any representation tried: the source analyses suggest that much of that signal reflects persistent company characteristics already captured by the hybrid, while scientific text provides an exploratory three-month increment motivating a matched comparison of text and structured event features. The next section separately examines hindsight in the frontier summary ratings and in prompted forecasts; it does not certify the embedding results, whose models' exposure is a matter of provenance (Table~\ref{tab:prov}): the pre-2020 model's pretraining closes in 2017, the current model's cutoff is undisclosed.

\section{Is It Hindsight? A Pre-2020 Model Against Current Models}
\label{sec:leak}

A rater trained on text through 2025 may have encountered the outcome of the trial a 2021 article describes. The screen already says that whatever such a rater knows did not reach the ranking, since no frontier polarity column is admitted; this section asks how much of its apparent skill as a rater reflects subsequent outcome information, which needs a comparison rater trained on pre-2020 data. The ledger's polarity finding rests on FinBERT (checkpoint \texttt{ProsusAI/finbert}; Table~\ref{tab:prov}), whose published training data predate the sample but which rates 38--52\% of biotech records neutral, a rate consistent with a domain mismatch though not proof of one (no vocabulary audit was run), and on Claude Sonnet 4, which knows the vocabulary and may know the outcomes. A rater that learned the domain's cues from pre-2020 text alone adds the control that separates the two explanations, though differences in model quality and domain fit mean the contrasts do not isolate hindsight by themselves.

\paragraph{A rater trained on pre-2020 data.} GPT-2 small \citep[124M parameters, released February 2019, pretrained on WebText, which excludes links created after December 2017;][]{radford2019gpt2} is fine-tuned as a three-class classifier on labels that cannot carry hindsight: 18{,}604 pre-2020 records from the system's own feeds, labelled by a deterministic rulebook that encodes the frontier chain's own materiality guide, together with the Financial PhraseBank \citep{malo2014good} and FiQA 2018 \citep{maia2018fiqa} (8{,}613 labelled rows, 7{,}751 training and 862 hold-out; Appendix~\ref{app:rater}). Hold-out accuracy is 88.4\% (762 of 862 rows; macro-F1 0.875); on the 53 pre-2020 StockTwits posts whose authors labelled them bullish or bearish, the rater's sign agrees with the author on 38 (72\%). It scores about 490k deduplicated raw records and the 62{,}492 daily summaries as $P(\text{pos})-P(\text{neg})$, aggregated and screened exactly as the FinBERT columns are. The twelve GPT-2-rated columns are ten raw-record scores (six per-source scores, two tier aggregates and their two 90-day means), one rating of the frontier-written summaries and one recency column; ``clean raw-record chain'' names the rater on raw records, clean end to end, while ``GPT-2 (summaries)'' is clean only as a rater, since those inputs inherit whatever the 2026 summarizer knew. Pre-specified (H1): no GPT-2-rated column is admitted in any fold--horizon pair. None is, in eighteen pairs; the closest miss is the summary rating at twelve months in the 2020 fold, $|t|$ 3.39 against 4, and the row is zero by construction with a post hoc magnitude bound of 0.063 at twelve months (H4 met). The absence of a measurable content gain does not depend on which clean rater reads it.

\paragraph{Design.} Four raters score identical text: the 62{,}492 GPT-5.4-mini daily summaries are rated by Claude Sonnet 4 (training data to 31 March 2025 \citep{anthropic2025claude4}), by GPT-5.4-mini itself under the same prompt (cutoff 31 August 2025 \citep{openai2026gpt54mini}), by FinBERT and by the GPT-2 rater. The split is target-end-after-cutoff: a formation month is post-cutoff at horizon $h$ if its target window ends after the rater's cutoff, so post windows are formed up to $h$ months before the cutoff and their returns partly precede it; at twelve months only two formations lie wholly after Claude's cutoff and none after GPT-5.4-mini's (counts in Table~\ref{tab:panels}). For a frontier--clean pair, a gap that falls across the cutoff is the proposed leakage signature; attributing it requires a stable clean--clean comparison, which the FinBERT-against-GPT-2 pair supplies as a placebo whose gap should not move, with the caveat that both raters share the Financial PhraseBank labels. Throughout, ``clean'' describes a rater's training data, which predate the evaluation sample; it says nothing about the pipeline that produced the inputs or about retrospective data coverage, which affects every column alike. On the pipeline, the summaries are written by GPT-5.4-mini, whose 31 August 2025 cutoff governs every row's inputs: the Claude and GPT-2 comparisons on summaries are rater-clean only, the GPT-5.4-mini comparison has a common summarizer--rater cutoff, and only the raw-record chain is clean end to end. The GPT-2 rater's fine-tuning records run to December 2019, so 24 formation months of this panel overlap its fine-tuning period (rule labels, not outcomes; the evaluated summaries are not necessarily training examples); those comparisons have training-period overlap, while the 2020--2025 ledger rows are cleanly separated (conventions in Appendix~\ref{app:cutoff}, provenance in Table~\ref{tab:prov}).

\begin{table}[!b]
\centering
\caption{Leakage on the rating task. Monthly rank IC of each rater's carried-forward rating of the same daily summaries with the realized excess return, on the months before and after the frontier rater's training cutoff; the paired frontier-minus-clean gap in each window; the before-minus-after difference with Newey--West $t$ and a block-bootstrap interval. Rows where both raters are clean are placebos. Target windows ending after the cutoff define the post months.}
\label{tab:leak}
{\footnotesize\setlength{\tabcolsep}{3pt}
\resizebox{\linewidth}{!}{\begin{tabular}{llllrrrrrrl}
\toprule
$h$ & Rater A (cutoff) & Rater B & Months pre/post & IC$_A$ pre & IC$_A$ post & IC$_B$ pre & IC$_B$ post & Gap pre & Gap post & Pre $-$ post ($t$; boot 95\%) \\
\midrule
1m & Claude Sonnet 4 (2025-03) & FinBERT & 86 / 15 & $+0.020$ & $-0.007$ & $+0.004$ & $-0.016$ & $+0.016$ & $+0.008$ & $+0.007$ (0.6; $[-0.022, +0.035]$) \\
1m & Claude Sonnet 4 (2025-03) & GPT-2 (summaries) & 86 / 15 & $+0.020$ & $-0.007$ & $+0.016$ & $-0.017$ & $+0.004$ & $+0.009$ & $-0.005$ ($-0.4$; $[-0.029, +0.019]$) \\
1m & GPT-5.4-mini (2025-08) & FinBERT & 91 / 10 & $+0.017$ & $+0.024$ & $+0.002$ & $-0.005$ & $+0.015$ & $+0.028$ & $-0.013$ ($-0.9$; $[-0.048, +0.022]$) \\
1m & GPT-5.4-mini (2025-08) & GPT-2 (summaries) & 91 / 10 & $+0.017$ & $+0.024$ & $+0.012$ & $-0.005$ & $+0.004$ & $+0.029$ & $-0.024$ ($-2.0$; $[-0.050, +0.001]$) \\
1m & GPT-5.4-mini (2025-08) & Claude Sonnet 4 & 91 / 10 & $+0.017$ & $+0.024$ & $+0.016$ & $+0.012$ & $+0.000$ & $+0.011$ & $-0.011$ ($-1.2$; $[-0.034, +0.014]$) \\
1m & \textit{Placebo:} FinBERT (split 2025-03) & GPT-2 (summaries) & 86 / 15 & $+0.004$ & $-0.016$ & $+0.016$ & $-0.017$ & $-0.012$ & $+0.001$ & $-0.013$ ($-0.9$; $[-0.040, +0.014]$) \\
\addlinespace
3m & Claude Sonnet 4 (2025-03) & FinBERT & 83 / 15 & $+0.010$ & $+0.007$ & $+0.003$ & $-0.021$ & $+0.007$ & $+0.028$ & $-0.022$ ($-1.5$; $[-0.052, +0.006]$) \\
3m & Claude Sonnet 4 (2025-03) & GPT-2 (summaries) & 83 / 15 & $+0.010$ & $+0.007$ & $+0.010$ & $-0.012$ & $+0.000$ & $+0.018$ & $-0.018$ ($-1.2$; $[-0.050, +0.009]$) \\
3m & GPT-5.4-mini (2025-08) & FinBERT & 89 / 9 & $+0.013$ & $+0.028$ & $+0.002$ & $-0.020$ & $+0.011$ & $+0.048$ & $-0.036$ ($-1.9$; $[-0.072, -0.004]$) \\
3m & GPT-5.4-mini (2025-08) & GPT-2 (summaries) & 89 / 9 & $+0.013$ & $+0.028$ & $+0.009$ & $-0.017$ & $+0.004$ & $+0.044$ & $-0.040$ ($-2.9$; $[-0.070, -0.008]$) \\
3m & \textit{Placebo:} FinBERT (split 2025-03) & GPT-2 (summaries) & 83 / 15 & $+0.003$ & $-0.021$ & $+0.010$ & $-0.012$ & $-0.006$ & $-0.010$ & $+0.003$ (0.2; $[-0.033, +0.037]$) \\
\addlinespace
12m & Claude Sonnet 4 (2025-03) & FinBERT & 74 / 15 & $+0.024$ & $+0.039$ & $+0.019$ & $-0.014$ & $+0.005$ & $+0.053$ & $-0.048$ ($-1.9$; $[-0.088, -0.011]$) \\
12m & Claude Sonnet 4 (2025-03) & GPT-2 (summaries) & 74 / 15 & $+0.024$ & $+0.039$ & $+0.004$ & $+0.041$ & $+0.020$ & $-0.002$ & $+0.022$ (1.4; $[-0.009, +0.051]$) \\
12m & GPT-5.4-mini (2025-08) & FinBERT & 79 / 10 & $+0.051$ & $+0.032$ & $+0.017$ & $-0.013$ & $+0.034$ & $+0.045$ & $-0.011$ ($-0.4$; $[-0.063, +0.037]$) \\
12m & GPT-5.4-mini (2025-08) & GPT-2 (summaries) & 79 / 10 & $+0.051$ & $+0.032$ & $+0.009$ & $+0.021$ & $+0.042$ & $+0.011$ & $+0.031$ (1.7; $[-0.001, +0.067]$) \\
12m & \textit{Placebo:} FinBERT (split 2025-03) & GPT-2 (summaries) & 74 / 15 & $+0.019$ & $-0.014$ & $+0.004$ & $+0.041$ & $+0.015$ & $-0.055$ & $+0.070$ (2.8; $[+0.030, +0.109]$) \\
\bottomrule
\end{tabular}}}
\par\medskip\begin{minipage}{\linewidth}\footnotesize\raggedright\setlength{\parskip}{1pt}
\par Ratings are carried forward within a company's observed span; months with fewer than twenty rated names are dropped. The split is target-end-after-cutoff (Appendix~\ref{app:cutoff}), so the post windows are nine to fifteen overlapping observations whose returns partly precede the cutoff; the GPT-5.4-mini rows are the comparison with a common summarizer--rater cutoff (the summarizer is the rater), the Claude rows rater-clean only. Bootstrap intervals resample each window in blocks of $h$ months. At three and twelve months the two frontier raters differ by $-0.003$ and $+0.011$ across the later cutoff; the placebo at that cutoff gives $-0.011$, $-0.004$ and $+0.042$. Figure~\ref{fig:cutoff} (Appendix~\ref{app:cutoff}) plots the one-month series with both cutoffs marked. H3 series, GPT-2 rater on summaries minus on raw records (Claude / GPT-5.4-mini splits): before the cutoffs $+0.016$ / $+0.013$ at one month ($t$ 1.5 / 1.2), $+0.006$ / $+0.003$ at three, $-0.031$ / $-0.017$ at twelve; after them $-0.020$ / $-0.012$, $-0.032$ / $-0.030$ and $+0.130$ / $+0.099$. The GPT-2 rater's raw-record IC in those cells: before, $-0.000$ / $-0.000$, $+0.004$ / $+0.006$, $+0.036$ / $+0.026$; after, $+0.004$ / $+0.007$, $+0.020$ / $+0.014$, $-0.089$ / $-0.078$.
\end{minipage}
\end{table}

\paragraph{Rating.} Table~\ref{tab:leak} gives the result for H2 against three outcomes written down in advance: a gap present before the cutoff that falls after it with a stable placebo (leakage); a gap small before the cutoff that does not fall (no edge for hindsight to explain); and a fall the placebo matches (unidentified). At one and three months the frontier-minus-GPT-2 gap is at most $+0.004$ before the cutoffs, small relative to the reported uncertainty, and it does not fall; the Claude--FinBERT one-month gap declines by $0.007$, with an interval, $[-0.022, +0.035]$, that includes zero and the whole pre-cutoff gap, and its gaps widen at three and twelve months because FinBERT's post-cutoff IC turns negative. The GPT-5.4-mini-against-GPT-2 comparison at three months widens ($-0.040$, $[-0.070, -0.008]$) for the same reason, the direction opposite to leakage, and is not interpreted. At twelve months, the one horizon with an edge to lose ($+0.020$ and $+0.042$), both gaps fall, by $+0.022$ and $+0.031$, the signature leakage would leave; but the clean--clean placebo falls by $+0.070$ ($t$ 2.8) at the March split and $+0.042$ at the August split, because GPT-2's twelve-month IC rises and FinBERT's drops after March 2025, so the parallel-trends assumption fails and the twelve-month contrasts do not identify contamination. The same windows answer H3: before the cutoffs the GPT-2 rater's rating of the frontier summaries leads its rating of the raw records by at most $+0.016$ ($t$ 1.5) and after them it trails, summary ratings lead raw-record ratings before the cutoffs and trail them afterward at one and three months, so the post-cutoff results support H3 directionally, while the pre--post changes have not been tested directly and do not identify contamination; the twelve-month cell is among the unstable ones. These comparisons do not establish absence of leakage; they show that whatever a frontier rater knows about 2020--2025 outcomes did not appear as a pre-cutoff rating edge that vanishes afterwards, consistent with the ledger: content carries little that ranks companies within a month, with or without subsequent outcome information.

\paragraph{Forecasting by prompt, identity-blind.} The second task asks the frontier model to forecast. For a fixed-seed sample of 300 companies and every evaluation month (15{,}627 company-months), the exact signal payload the ranking uses is serialized as text: the price-model score's percentile within the month, eight price and volume features, eight balance-sheet fields, five attention counts and the latest anonymized daily summary (Appendix~\ref{app:prompts}). GPT-5.4-mini is told to use only the supplied signals, to use no outside knowledge about any company, drug, market or period, and to return one integer from 0 to 100, its estimate of the company's rank by twelve-month excess return; the same twelve-month score is evaluated against returns at all three horizons, so the one- and three-month rows measure cross-horizon association against horizon-specific comparators. Each payload is sent twice: \emph{blind}, with the ticker, the company name, the date and drug codes masked, which removes the explicit identifiers but not every cue from which the company might be inferred; and \emph{revealed}, with them included. The revealed-minus-blind contrast measures the contribution of explicit identifiers and dates; a larger pre-cutoff value is consistent with recall despite the instruction against outside knowledge, with the caveat that post-cutoff targets partly precede the cutoff.

Table~\ref{tab:prompt} (Appendix~\ref{app:cutoff}) gives the result for H7. Blind, the frontier model forecasts about as well as the one line that summarizes the price history: its twelve-month neutral IC of $0.078$ sits below the market score's $0.096$ and the hybrid's $0.183$ on the same rows, and blending it into the hybrid is given zero weight in every fold. Revealing the company and the month helps, and helps more when the outcome was knowable: the revealed score beats the blind one by $+0.025$ IC at twelve months and $+0.013$ at three before the cutoff, and by $+0.008$ and $+0.001$ after it; the before-minus-after difference is $+0.017$ ($t$ 1.7) and $+0.012$ ($t$ 2.5). That is the signature the rating task did not show, about a fifth of the blind model's raw IC ($+0.025$ of 0.134), and it appears although the prompt forbade outside knowledge: an instruction not to recall is not a control for recall. The post-cutoff windows are five to ten months (Table~\ref{tab:panels}).

\paragraph{The narrative layer, checked mechanically.} The descriptive layer writes, for each company and horizon, a JSON narrative from the same inputs the ranking uses, under a rulebook with two defining rules (describe, do not forecast; use only the supplied signals) and six rules of form. The system's own audit of this layer uses a frontier model, GPT-5 \citep{openai2025gpt5}, as judge. Here every one of the 300 narratives (100 companies $\times$ three horizons, queried as of 30 June 2025) is checked by six deterministic rules (Appendix~\ref{app:faith}): numbers and entities must trace to the input, categories must be input fields, the forecast lexicon must be absent, nulls must be honoured and no evaluative adjective may attach to a number. Table~\ref{tab:faith} reports the result beside the judge's verdicts.

\begin{table}[!htbp]
\centering
\caption{Mechanical compliance of the narrative layer, 300 narratives (100 companies $\times$ three horizons): share passing each deterministic rule, and the agreement of the frontier-model judge (GPT-5) with the rules on the outside-knowledge rule it was asked to enforce.}
\label{tab:faith}
{\footnotesize\setlength{\tabcolsep}{4pt}
\begin{tabular}{lrrrrrrr}
\toprule
Horizon & Numbers & Entities & Categories & No forecast & Nulls & No evaluative & All \\
 & trace & trace & supplied & lexicon & honoured & words & rules \\
\midrule
1 month & 100/100 & 100/100 & 72/100 & 100/100 & 100/100 & 100/100 & 72/100 \\
3 months & 100/100 & 100/100 & 75/100 & 100/100 & 100/100 & 100/100 & 75/100 \\
12 months & 100/100 & 100/100 & 74/100 & 100/100 & 100/100 & 100/100 & 74/100 \\
\midrule
\multicolumn{8}{l}{Judge against rules, outside-knowledge rule, 299 judged narratives: rules fail and judge flags, 14; rules fail and judge passes, 65;} \\
\multicolumn{8}{l}{rules pass and judge flags, 0; rules pass and judge passes, 220. Agreement 0.78, Cohen's $\kappa$ 0.24.} \\
\bottomrule
\end{tabular}}
\par\medskip\begin{minipage}{\linewidth}\footnotesize\raggedright\setlength{\parskip}{1pt}
\par 6{,}252 numeric tokens, every one an exact match to an input value (the generator prints floats verbatim); 1{,}599 null fields, none given a value; zero forecast-lexicon hits (three occurrences of ``do not constitute a forecast'' are negated mentions). The single recurring failure is the word ``regulatory'' in ``scientific and regulatory event data is unavailable'', a category that is not a field of the input schema, in 79 narratives. The judge flagged 14 of those 79 and none of the 220 rule-passing narratives it judged; one three-month narrative (TKNO) failed the generator's schema validation and was not judged.
\end{minipage}
\end{table}

All 300 narratives pass five of the six specified checks, and a supplementary scan for the five forecast terms the prompt forbids but the checker's lexicon omitted (could, may, on track, outlook, risk of) finds none; 79 fail the category check on one fixed phrasing. The frontier judge, asked to enforce a rule it could check by inspection, flagged 14 of those 79 and none of the 220 rule-passing narratives, so its 4.7\% flag rate understated a 26\% rate; the judge results measure detection of this one violation, not general factual accuracy. The checks do not exhaust the prompt's requirements (the missing-field rule is tested as no value supplied, not as the exact phrase), and no human semantic evaluation was conducted, so semantic faithfulness is not established. Lift is not claimed for a describe-only layer; the pre-specified prescience test (narrative tone against realized returns across the generator's cutoff) was not run, and faithfulness does not bound it, since emphasis among supplied facts remains a channel. Generation dates and prompt provenance are in Table~\ref{tab:prov} and Appendix~\ref{app:prompts}.

\section{When Each Source Works}
\label{sec:state}

The gains the ledger establishes come from numbers and counts, and the text results of Sections~\ref{sec:clean} and~\ref{sec:leak} leave them as the ranking's information. The last question is when each works. The two channels that carry the ledger's gain alternate within the sample. Fundamentals carry the twelve-month gain in 2024, a year that followed a sector rally, and in 2020, where attention had no admitted column yet (Table~\ref{tab:years}, Appendix~\ref{app:tables}); attention carries it in 2022 and 2025, years that followed a decline; the gain as a whole is largest in the wide-credit-spread years 2020, 2022 and 2023. A circular block permutation of the state labels (twelve-month blocks, 10{,}000 draws) gives unadjusted evidence of state dependence for the fundamental-minus-attention difference, a paired series on the 53 months from 2021 in which both channels have admitted columns, with terciles re-cut on those months: $+0.049$ across terciles of trailing twelve-month XBI return ($p=0.023$) and $+0.054$ across terciles of realized volatility ($p=0.011$). The two channels' own high-minus-low contrasts on all 65 months are $+0.034$ and $-0.031$ (trailing return) and $+0.046$ and $-0.025$ (volatility), each with its own permutation $p$ (Table~\ref{tab:stategrid}); the paired difference is not their arithmetic difference because its sample and terciles differ. Across terciles of the high-yield spread, the full-model gain differs by $+0.039$ ($p=0.039$). These splits were specified after the year-by-year results were seen; six years hold a small number of distinct episodes of each kind (Table~\ref{tab:states}, Appendix~\ref{app:tables}, gives the state-month composition and the permutation's resolution). This pattern uses numerical and attention features rather than language-model scores; it remains subject to retrospective coverage and exploratory state selection. A plausible reading, offered as such: after a rally and in volatile months, investors re-price balance sheets, so a company's liabilities and cash position discriminate; after a decline, selection runs on which companies still draw coverage, so attention discriminates. It is a hypothesis for later data, not a finding.

\paragraph{A constructive test.} We test whether state-dependent weights improve the two-channel blend. For each held-out year $Y$ a blend $b_w = w\,\mathrm{rank}_{\text{fund}} + (1-w)\,\mathrm{rank}_{\text{att}}$ of the two channel models' within-month ranks (the fundamental family model and the sentiment-pipeline family model, whose admitted columns are all attention counts, so it is the attention channel of Table~\ref{tab:ledger}) is fitted on the other five years, with a separate $w$ for each tercile of the state at formation, chosen to maximize the neutral IC on the training months in that tercile (grid $w\in\{0,0.1,\ldots,1\}$), and applied to year $Y$. The comparator is the same blend with one $w$ fitted without state. The design is semi-out-of-sample, because the state terciles are cut on all 65 months and the two states were chosen after the pattern was seen; it is leave-one-year-out in the weights, and those weights are retrospective: they are fitted on the other evaluation years, not on data that precede the test year as a prospective walk-forward rule would be. Specified before the blend was run, after the pattern had been seen (H5): the regime blend's twelve-month neutral IC exceeds the static blend's.

\paragraph{Every channel on the same grid.} The same tercile test applied to every channel gain, every polarity column and the first text panel's head and identity placebo (Table~\ref{tab:stategrid}, Appendix~\ref{app:tables}) sorts the signals into two groups: the fundamental gain, the frozen-text head and its identity placebo concentrate after rallies and in volatile months, attention is the mirror image, and no polarity column in Table~\ref{tab:stategrid} has state structure at the 5\% level on the two pre-specified twelve-month states except the GPT-2 social score. The text head therefore behaves like a balance-sheet signal, which is what a persistent company-level signal should do, and not like attention.

\paragraph{The proposed model, and whether text belongs in it.} The natural final model is the regime blend extended to three rankings: market-only, the hybrid and the LLM text ranking of Section~\ref{sec:clean}, with the three weights chosen per state tercile on the other five years (66 weight triples, step 0.1) and applied to the held-out year, rows without text taking the hybrid's rank; Table~\ref{tab:threeway} (Appendix~\ref{app:tables}) gives every cell. Pre-specified (H8): the blend's twelve-month neutral IC exceeds the hybrid's. The recorded rule names the direction and no significance threshold, so the verdict applies the paper's reporting convention, a two-sided 5\% Newey--West test ($|t| \geq 1.96$) on the paired twelve-month difference: no cell meets it, for either text model or for scientific text, although point estimates exceed the hybrid in several cells. The twelve-month evidence is mixed rather than uniformly flat: the high-yield-state scientific-text variant has a positive neutral bootstrap interval, $[+0.001, +0.024]$, with Newey--West $t$ 1.7 and a raw-IC decline of 0.031. At three months the scientific-text blend on covered rows carries the same lead in regime form ($+0.006$ to $+0.009$ over the hybrid), in no state larger than the static weight gives, so state conditioning does not establish an advantage over the static text blend. Coverage matters: the scientific-text static three-way blend at three months gains $+0.017$ neutral IC across all rows but loses $0.024$ raw IC against the hybrid. The all-row implementation changes covered companies' positions relative to uncovered ones (a covered company's text percentile is computed among the companies with text that month, an uncovered company takes its hybrid percentile over the whole month, and the blend does not slot-preserve), so its neutral gain combines text-score changes with cross-group reordering while raw IC declines; coverage correlates with size and liquidity, which the neutral metric adjusts for and the raw metric does not, so the covered-sample result does not establish improved raw ranking across the full universe. The model this paper proposes is therefore the hybrid, with the attention counts coming from the data pipeline rather than from any language model and the three-month scientific-text overlay as the one addition to test on later data (Section~\ref{sec:conclusion}).

H5 is met on both states named in advance (Table~\ref{tab:regime}, Appendix~\ref{app:tables}): the regime blend adds $+0.013$ neutral IC over the static blend, with intervals that exclude zero, and the fitted weights follow the observed state pattern, fundamentals alone in the high tercile (after rallies, in volatile months), attention alone in the low tercile and a mixed weight in the middle, in every fold. Two qualifications bound the result. Exploratory state-dependent blends outperform the static two-channel blend in this sample; their advantage over the full hybrid ($+0.004$ to $+0.006$, $t$ 0.6 to 0.9) is not statistically established, so the state structure is an interpretable route to information the full hybrid already holds; and conditioning on the credit spread, the state that moves the level of the gain rather than its composition, hurts. The raw gain is $+0.004$ to $+0.005$, with $t$ 1.8 to 2.0.

\section{Related Work}
\label{sec:related}

Machine learning for the cross-section of returns \citep{gu2020empirical}, the multiple-testing problem of factor discovery \citep{harvey2016cross} and the finding that most published factors nonetheless replicate \citep{jensen2023replication} frame the screen; media sentiment, news representations and language-model signals \citep{tetlock2007giving, ke2019predicting, lopezlira2023chatgpt, chen2023expected} frame the text tests, and the finding that LLM-based timing strategies' advantages shrink in broader, longer backtests \citep{li2026llm} is the closest precedent for the text result, though it concerns timing strategies rather than cross-sectional ranking signals. \citet{glasserman2023lookahead} compare original and anonymized headlines inside and outside the model's training window, \citet{sarkar2024lookahead} test for information unavailable at the historical prediction date, and \citet{ludwig2025llm} treat look-ahead as an identification problem for econometric use; the design here holds the text fixed and varies the rater, so the same article is scored by models that could and could not know its sequel, with a clean--clean placebo. FinBERT \citep{araci2019finbert}, the Financial PhraseBank \citep{malo2014good} and GPT-2 \citep{radford2019gpt2} supply the cutoff-clean baseline rater, the labels and a pretrained model whose data window closes before the sample.

\section{Limits and Conclusion}
\label{sec:conclusion}

Every non-market column exists only for companies listed in June 2026, so every gain in the ledger is a within-survivor estimate of unknown bias sign, and attention's recency count in particular may reflect retrieval-time coverage; without it the one-, three- and twelve-month gains are zero, $+0.009$ and $+0.022$ (Table~\ref{tab:ledger}, note). Twelve-month claims rest on 65 overlapping months and carry bootstrap intervals. The post-cutoff windows are nine to fifteen months on a thin, late-sample slice of company-days, so the leakage estimates are wide, and the twelve-month contrasts, where the two clean raters themselves diverge, do not identify contamination; a test with power needs target windows that lie wholly after the cutoff, which the data will supply as 2026 and 2027 resolve. The state results are exploratory in their specification and semi-out-of-sample in their test. The GPT-2 rater is a 124M-parameter model fine-tuned on headline-length weak labels; a stronger pre-2020 rater could change the GPT-2-rated row. The narratives are checked by deterministic rules for faithfulness, not by human readers and not for usefulness; the prescience test was specified and not run.

Within those limits the ledger supports incremental ranking evidence on a survivor sample: the IC is a ranking association, and no return, cost or implementation claim is made. A machine-learned healthcare ranking learns from price trends first, then from balance sheets and from how much attention a company receives, and little from what the text says beyond persistent company-level language, with one three-month exception. The text rankings show twelve-month predictive association, much of it consistent with persistent company information already captured by the numbers; scientific text provides an exploratory three-month increment on covered rows, with a post hoc event-month result motivating further testing. The matched-text comparisons show negligible pre-cutoff frontier-minus-GPT-2 gaps at one and three months that do not shrink after the cutoffs, and do not identify hindsight bias at twelve; no language-model score enters the recommended hybrid, which removes that direct contamination channel while leaving retrospective coverage unresolved, and the scientific-text embedding model has an undisclosed cutoff. In prompted forecasting, the revealed-minus-blind gap falls across the cutoff at every horizon, nominally significant at three months. The narrative layer passes every mechanical check but one boilerplate category violation, which the frontier judge missed four times in five; mechanical compliance does not establish semantic faithfulness or usefulness to analysts, and no human evaluation was conducted. The retrospective regime blends describe an interpretable state pattern, with one of the two fundamentals years (2020) holding by construction, without establishing an advantage over the full hybrid or prospective performance. The hybrid remains the reference ranking: a market score plus screened balance-sheet and attention signals, with no language model inside it. Scientific-text augmentation at three months is a candidate for testing on unused data, with a matched text-versus-structured comparison on the same rows, an identity-only baseline and wholly post-cutoff windows as the tests that decide it.

\paragraph{Code and data availability.} No public repository accompanies this submission. The analysis code, the verbatim prompts (Appendix~\ref{app:prompts}), the text-model settings (Appendix~\ref{app:tables}) and the derived per-company-day scores with cutoff flags can be provided on request, subject to the authors' institutional compliance review; the licensed source data (prices and returns, FactSet classifications, news, social and analyst feeds, filings and registry extracts) cannot be redistributed. Every number in the paper traces to a saved output of that code.

\bibliographystyle{ACM-Reference-Format}
\bibliography{references}

\clearpage
\appendix
\thispagestyle{plain}

\section{Data Volumes, Screen Rule and Admissions}
\label{app:data}

\begin{table}[!htbp]
\centering
\caption{Data volume by source pipeline, January 2010 to June 2026 (raw, pre-screen counts over the unscreened universe), and the non-market columns admitted by the $|t|\geq4$ screen in at least one of the six test-year folds, with admitting folds out of six by horizon and the range of the admission $t$.}
\label{tab:volumes}
{\footnotesize
\resizebox{\linewidth}{!}{\begin{tabular}{lrrrr}
\toprule
Pipeline & Source records & Firm-months & Companies & Columns \\
\midrule
Market & 10{,}130{,}326 & 486{,}174 & 4{,}868 & 14 \\
Sentiment (news, social, analyst tiers) & 2{,}097{,}935 & 88{,}032 & 1{,}010 & 45 (42 distinct) \\
Fundamental (XBRL) & 33{,}309 & 92{,}912 & 870 & 19 \\
Scientific (trials, FDA, releases, labels, PubMed, bioRxiv) & 24{,}680 & 110{,}366 & 1{,}001 & 123 \\
Macroeconomic (VIX, HY spread, XBI drawdown) & 4{,}289 & 198 & -- & 13 (3 screened) \\
\midrule
\multicolumn{5}{l}{\textit{Admitted columns (folds of six at 1m / 3m / 12m; admission $t$)}} \\
\multicolumn{5}{l}{Sentiment-pipeline record recency (attention) 6/4/4, $-5.35$ to $-4.18$; analysts covering the stock (attention) --/--/5, 4.69 to 7.85} \\
\multicolumn{5}{l}{Total liabilities --/4/6, 4.20 to 6.47; days since last filing --/3/2; revenue --/--/2; financing risk --/--/1; total assets --/--/1} \\
\multicolumn{5}{l}{Press-release recency --/--/3, $-4.78$ to $-4.25$; clinical-trial recency --/--/1} \\
\multicolumn{5}{l}{Never admitted: every text-polarity column (frontier chain, closest $|t|$ 3.96), every analyst rating (3.67), every macro conditioning variable, every GPT-2-rated column (closest $|t|$ 3.39)} \\
\bottomrule
\end{tabular}}}
\par\medskip\begin{minipage}{\linewidth}\footnotesize\raggedright\setlength{\parskip}{1pt}
\par Text sources read by the text tower (Section~\ref{sec:clean}): news (Finnhub titles and snippets, GDELT and Google News titles), social posts (Reddit, Yahoo boards, StockTwits where fetched) and scientific text (SEC press-release headlines, clinical-trial registry entries, PubMed and bioRxiv titles, FDA event and advisory-committee text). Outcome coverage of the 2020--2025 evaluation: 59{,}364 firm-months over 72 months at one and three months, 53{,}963 over 66 at twelve (65 with at least twenty names); exits are completed at the last observed price plus XBI exposure (0.5\%, 1.4\%, 4.8\% of rows). The screen tests 184 distinct firm-level columns plus three macro conditioning variables per fold, 561 tests; the within-fold 5\% Bonferroni value is 3.92. Companies of the three non-market families are survivors as of June 2026 (Section~\ref{sec:conclusion}).
\end{minipage}
\end{table}

\section{Prompts and the Rule Labeler}
\label{app:prompts}
\paragraph{P3, daily news synthesis (GPT-5.4-mini).}
\begin{lstlisting}[style=prompt]
You are a biotech/healthcare financial news classifier.
You extract structured metadata from news articles for use in a quantitative investment model.
You MUST respond with valid JSON only - no prose, no markdown, no explanation, no code fences.
Any deviation from the schema will break a downstream pipeline.

Synthesize cross-source news for {ticker} on {date} into a single investment-grade daily narrative.

Sources (Finnhub / GDELT / Google News / SEC filings):
{text}

Return ONLY a JSON object matching this exact schema:
{
  "news_summary": "<1-2 sentences synthesizing across sources: state the dominant event and outcome, name the drug/trial/asset involved, include any quantitative details (deal size, trial phase, price target), and state the net investment implication (catalyst, risk, or sentiment shift)>",
  "news_themes": ["<1-3 items from this list ONLY: ['earnings', 'regulatory', 'product', 'partnership', 'management', 'macro', 'analyst', 'ai_tech', 'clinical', 'financing', 'operations', 'other']>"],
  "news_materiality": "<MUST be exactly one of: high | medium | low>"
}

Materiality guide:
- high: FDA decision, phase 3 data readout, M&A, major partnership, bankruptcy
- medium: phase 1/2 data, analyst rating change, earnings beat/miss, financing round
- low: conference presentation, executive quote, minor pipeline update, general market commentary
\end{lstlisting}
\noindent\begin{minipage}{\linewidth}
\paragraph{P6, rating of the daily summary (Claude Sonnet 4 and GPT-5.4-mini, temperature 0).}
\begin{lstlisting}[style=prompt]
Rate the sentiment of this text SPECIFICALLY for the stock ticker {ticker} on a scale from -1.0 (very bearish) to +1.0 (very bullish). Ignore sentiment about other companies. Return ONLY a single number between -1.0 and 1.0.

Text: {text}

Score:
\end{lstlisting}
\end{minipage}
\paragraph{Narrative generator and judge (Section~\ref{sec:leak}).} The generator is Claude Sonnet 4 at temperature 0 under a fixed system prompt whose defining rules read, verbatim: ``1. Describe, do not forecast. No forward-looking language: not `will', `expected', `likely', `should', `poised', `on track', `could', `may', `outlook', `upside', `downside', `risk of'. [\ldots] 2. Use only the supplied values. No knowledge of the company, its drugs, trials, management, competitors, sector, or any event from outside the JSON. [\ldots] 3. Missing fields: when a field or block is null, write `no value is supplied for <field name>' using the field's own name, once, and move on. Do not introduce a category that is not a field.'' The six rules of form that follow (length, structure, tone) are not shown. The judge is GPT-5 at temperature 0 under a system prompt that restates the two defining rules as RULE 1 (describe, do not forecast) and RULE 2 (use only the supplied signals, with four violation classes: an outside fact or name, a category that is not a field, a null described otherwise, a number not in the input) and asks for JSON with pass flags and quoted violating spans. Both prompts were written in June 2026; their full texts were archived from the generation and audit notebooks on 27 September 2026 and are available with the code. Generator and judge are different vendors' models, but both prompts were written by the authors, so their independence is that of two models, not of two evaluators.

\paragraph{Execution details.} Synthesis (P3) ran at temperature 0.1 in June 2026; failed calls were retried, and the retry count was not logged. The GPT-5.4-mini re-rating (P6) made up to five attempts per call with exponential backoff on rate-limit and server errors, re-asked once with a longer token budget when the output was truncated, and left two of 62{,}492 rows unrated. Prompted forecasting made up to six attempts per call with backoff, then a second pass at lower concurrency over the rows still unparsed, after which 18 blind and 20 revealed rows of 15{,}627 remained unparsed and are dropped. The forecasting prompt text was fixed when the experiment was specified; two pilot runs changed the token budget (8 to 600) and set low reasoning effort, nothing else. The number of prompt versions tried in the production pipeline for P3 and P6 was not logged. Dates on bibliography entries are documentation retrieval dates, not inference dates; inference dates are in Table~\ref{tab:prov}.

\paragraph{Prompted forecasting (Section~\ref{sec:leak}), system message.}
\begin{lstlisting}[style=prompt]
You are scoring one US healthcare or life-science company at one month end for a quantitative ranking study. Use ONLY the signals supplied in the message. Do not use any outside knowledge about any company, drug, market or time period, and do not try to work out which company or which month this is. Output a single integer from 0 to 100: your estimate of where this company will rank among its peers by 12-month excess return over the XBI index, where 100 means the top of the cross-section and 0 the bottom. Return only the integer, with no words, units or explanation.
\end{lstlisting}
The user message lists the signals as ``label: value'' lines (missing values as ``not supplied'') and, for the blind arm, no ticker, name or date; the summary text has the ticker, the company's name tokens, exchange tags and drug or program codes replaced. The revealed arm prepends ``Ticker / Company / Month'' lines.

\paragraph{Rule labeler.} Two case-insensitive regular-expression lexicons over the record text (title, or title plus snippet or body); a record matching only the positive lexicon is labelled positive, only the negative lexicon negative, neither neutral, both dropped. The positive lexicon has 34 patterns (approvals, met endpoints, designations, deals, upgrades, beats, patents), the negative 46 (complete response letters, failed or halted trials, clinical holds, going concern, offerings and dilution, downgrades, misses, recalls, bankruptcy, delisting, litigation, safety signals, guidance cuts); both lexicons are available with the code. Applied to the 18{,}604 pre-2020 records (Reddit 8{,}195, Google News 6{,}378, GDELT 4{,}016, Finnhub 15) it labels 1{,}932 positive, 692 negative and 15{,}898 neutral and drops 82; neutral is downsampled to 2{,}624 so that it is at most half of the rule-labelled set. Public pre-2020 sets (fetched counts; 2{,}259 and 1{,}106 rows after deduplication, Table~\ref{tab:rater}): Financial PhraseBank, all-annotators-agree subset, 2{,}264 sentences \citep{malo2014good}, loaded as the Hugging Face dataset \texttt{takala/financial\_phrasebank}, configuration \texttt{sentences\_allagree}; FiQA 2018 headlines and posts, 1{,}213, loaded as \texttt{pauri32/fiqa-2018}, with the continuous score cut at $\pm0.1$ (both fetched with \texttt{datasets} 2.21.0 on 1--2 October 2026; the papers establish the datasets, these copies are what the counts describe); SemEval-2017 Task 5 was not retrievable and is not used. 8{,}613 labelled rows: 7{,}751 training and an 862-row hold-out (10\%, stratified by label). In a read of 45 polarity labels and 15 neutrals from a fixed-seed sample of 200 rule-labelled records, 35 polarity labels were clearly right, 5 weak, 3 wrong (the ticker CRL read as a complete response letter, ``DPD Beat'' read as an earnings beat, SEC approval read as a drug approval) and 2 about another company with the right polarity, and 13 of 15 neutrals were right. The labels are weak by design; the fine-tuned model is evaluated on the hold-out and on the external StockTwits check, not on the rules.

\section{Clean Rater: Training and Evaluation}
\label{app:rater}

\begin{table}[!htbp]
\centering
\caption{The clean rater: training labels by source and hold-out performance (10\% stratified by label, 862 rows). Base model \texttt{gpt2} (124M), sequence-classification head, three classes, two epochs, learning rate $2\times10^{-5}$, batch 16, maximum length 64 tokens for headlines and 128 for posts and summaries, seed 0; trained on serverless CPU (about 1.8 hours per epoch).}
\label{tab:rater}
{\footnotesize
\begin{tabular}{lrrrrrr}
\toprule
Label source & Negative & Neutral & Positive & Total & Hold-out accuracy & Macro-F1 \\
\midrule
Rules on Reddit posts (2010--2019) & 281 & 1{,}115 & 1{,}087 & 2{,}483 & 81.4\% & 0.75 \\
Rules on Google News headlines (to 2019) & 233 & 930 & 632 & 1{,}795 & 91.3\% & 0.91 \\
Rules on GDELT headlines (2017--2019) & 177 & 577 & 212 & 966 & 87.5\% & 0.86 \\
Rules on Finnhub (pre-2020) & 1 & 2 & 1 & 4 & -- & -- \\
Financial PhraseBank, all agree & 303 & 1{,}386 & 570 & 2{,}259 & 98.3\% & 0.98 \\
FiQA 2018 (score cut at $\pm0.1$) & 342 & 84 & 680 & 1{,}106 & 79.4\% & 0.68 \\
\midrule
All & 1{,}337 & 4{,}094 & 3{,}182 & 8{,}613 & 88.4\% & 0.875 \\
\bottomrule
\end{tabular}}
\par\medskip\begin{minipage}{\linewidth}\footnotesize\raggedright\setlength{\parskip}{1pt}
\par Hold-out per class: precision 0.83 / 0.94 / 0.85 and recall 0.85 / 0.85 / 0.94 for negative / neutral / positive. External check on the 53 pre-2020 StockTwits posts carrying an author label (39 bullish, 14 bearish): sign agreement 71.7\%, three-class accuracy 69.8\%, mean score 0.53 on bullish and 0.34 on bearish posts. Scoring: $P(\text{pos})-P(\text{neg})$ per record, averaged to company-day by source, with news and social tier means and the daily-summary score . Records scored: Finnhub 107,665, GDELT 15,967, Google News 65,905, Reddit 49,100, StockTwits 81,646, Yahoo 169,498 (posts truncated to 64 tokens at scoring), and 62,492 daily summaries (128 tokens); 204,255 company-days carry at least one score. On the summaries the clean rater agrees with FinBERT at Spearman 0.68 and with Claude Sonnet 4 at 0.61 (Claude and FinBERT agree at 0.68); 27\% of its summary scores lie within 0.1 of zero, against 32\% exactly zero for GPT-5.4-mini and 38--52\% neutral for FinBERT on raw records.
\end{minipage}
\end{table}

\section{Cutoff Test: Conventions}
\label{app:cutoff}

\begin{figure}[!htbp]
\centering
\begin{tikzpicture}
\begin{axis}[width=\textwidth, height=5.4cm, xmin=2018.9, xmax=2026.5, ymin=-0.16, ymax=0.16,
  xtick={2019,2020,2021,2022,2023,2024,2025,2026}, xticklabels={2019,2020,2021,2022,2023,2024,2025,2026}, tick label style={font=\scriptsize}, ytick={-0.1,0,0.1}, yticklabels={$-0.1$,0,0.1},
  ylabel={Rank IC, 1m, 3-month moving average}, xlabel={Formation month}, label style={font=\scriptsize},
  axis lines=left, axis line style={gray!60}, clip=true, legend style={at={(0.5,1.02)}, anchor=south, legend columns=4, draw=none, font=\scriptsize}]
\addplot[gray!70, line width=0.4pt, forget plot] coordinates {(2018.9,0) (2026.5,0)};
\addplot[gray!50, dashed, forget plot] coordinates {(2025.208,-0.16) (2025.208,0.16)};
\node[font=\tiny, gray, anchor=south west, rotate=90] at (axis cs:2025.208,-0.155) {Claude cutoff};
\addplot[gray!50, dashed, forget plot] coordinates {(2025.625,-0.16) (2025.625,0.16)};
\node[font=\tiny, gray, anchor=south west, rotate=90] at (axis cs:2025.625,-0.155) {GPT-5.4-mini cutoff};
\addplot[black, line width=0.9pt] coordinates {(2018.0417,0.1563) (2018.1250,0.0716) (2018.2083,0.0203) (2018.2917,0.0243) (2018.3750,0.0570) (2018.4583,0.1284) (2018.5417,0.0963) (2018.6250,0.1066) (2018.7083,0.0603) (2018.7917,0.0305) (2018.8750,0.0305) (2018.9583,0.0062) (2019.0417,-0.0050) (2019.1250,-0.0632) (2019.2083,-0.0122) (2019.2917,-0.0116) (2019.3750,-0.0339) (2019.4583,-0.0287) (2019.5417,-0.0199) (2019.6250,0.0500) (2019.7083,-0.0170) (2019.7917,-0.0248) (2019.8750,-0.0178) (2019.9583,-0.0059) (2020.0417,0.0186) (2020.1250,0.0369) (2020.2083,0.0473) (2020.2917,0.0223) (2020.3750,0.0064) (2020.4583,0.0242) (2020.5417,0.0452) (2020.6250,0.0964) (2020.7083,0.0428) (2020.7917,0.0486) (2020.8750,-0.0156) (2020.9583,-0.0137) (2021.0417,-0.0433) (2021.1250,-0.0453) (2021.2083,0.0099) (2021.2917,0.0137) (2021.3750,0.0457) (2021.4583,0.0475) (2021.5417,0.0610) (2021.6250,0.0432) (2021.7083,0.0216) (2021.7917,0.0070) (2021.8750,0.0082) (2021.9583,0.0157) (2022.0417,0.0295) (2022.1250,0.0447) (2022.2083,0.0226) (2022.2917,0.0034) (2022.3750,-0.0478) (2022.4583,-0.0127) (2022.5417,-0.0222) (2022.6250,0.0054) (2022.7083,-0.0141) (2022.7917,0.0070) (2022.8750,0.0106) (2022.9583,-0.0027) (2023.0417,0.0052) (2023.1250,0.0030) (2023.2083,0.0113) (2023.2917,-0.0183) (2023.3750,-0.0009) (2023.4583,0.0574) (2023.5417,0.0991) (2023.6250,0.0477) (2023.7083,-0.0002) (2023.7917,-0.0579) (2023.8750,-0.0378) (2023.9583,0.0087) (2024.0417,0.0417) (2024.1250,0.0495) (2024.2083,0.0102) (2024.2917,0.0299) (2024.3750,0.0406) (2024.4583,0.0745) (2024.5417,0.0574) (2024.6250,0.0604) (2024.7083,0.0165) (2024.7917,0.0239) (2024.8750,0.0061) (2024.9583,-0.0160) (2025.0417,-0.0003) (2025.1250,-0.0109) (2025.2083,0.0188) (2025.2917,-0.0373) (2025.3750,-0.0188) (2025.4583,-0.0700) (2025.5417,-0.0520) (2025.6250,-0.0406) (2025.7083,0.0177) (2025.7917,0.0365) (2025.8750,0.0167) (2025.9583,0.0069) (2026.0417,-0.0054) (2026.1250,0.0163) (2026.2083,0.0154) (2026.2917,0.0218) (2026.3750,0.0127)};
\addlegendentry{Claude Sonnet 4 (frontier)}
\addplot[black, line width=0.9pt, dashed] coordinates {(2018.0417,0.1118) (2018.1250,0.0128) (2018.2083,-0.0201) (2018.2917,0.0196) (2018.3750,0.0837) (2018.4583,0.1447) (2018.5417,0.1018) (2018.6250,0.1068) (2018.7083,0.0676) (2018.7917,0.0202) (2018.8750,0.0278) (2018.9583,0.0020) (2019.0417,0.0312) (2019.1250,-0.0154) (2019.2083,0.0087) (2019.2917,-0.0428) (2019.3750,-0.0690) (2019.4583,-0.0535) (2019.5417,0.0127) (2019.6250,0.0851) (2019.7083,0.0103) (2019.7917,-0.0324) (2019.8750,-0.0294) (2019.9583,0.0015) (2020.0417,0.0283) (2020.1250,0.0355) (2020.2083,0.0446) (2020.2917,0.0083) (2020.3750,-0.0151) (2020.4583,0.0024) (2020.5417,0.0266) (2020.6250,0.0785) (2020.7083,0.0196) (2020.7917,-0.0049) (2020.8750,-0.0832) (2020.9583,-0.0835) (2021.0417,-0.0864) (2021.1250,-0.0650) (2021.2083,0.0225) (2021.2917,0.0393) (2021.3750,0.0598) (2021.4583,0.0566) (2021.5417,0.0635) (2021.6250,0.0648) (2021.7083,0.0205) (2021.7917,0.0186) (2021.8750,-0.0031) (2021.9583,0.0101) (2022.0417,0.0212) (2022.1250,0.0415) (2022.2083,0.0332) (2022.2917,0.0117) (2022.3750,-0.0519) (2022.4583,-0.0252) (2022.5417,-0.0450) (2022.6250,0.0139) (2022.7083,-0.0050) (2022.7917,0.0467) (2022.8750,0.0517) (2022.9583,0.0426) (2023.0417,0.0127) (2023.1250,0.0243) (2023.2083,0.0295) (2023.2917,0.0201) (2023.3750,0.0009) (2023.4583,0.0375) (2023.5417,0.0827) (2023.6250,0.0576) (2023.7083,0.0222) (2023.7917,-0.0572) (2023.8750,-0.0512) (2023.9583,0.0042) (2024.0417,0.0361) (2024.1250,0.0617) (2024.2083,0.0312) (2024.2917,0.0786) (2024.3750,0.0957) (2024.4583,0.1159) (2024.5417,0.0988) (2024.6250,0.0577) (2024.7083,0.0068) (2024.7917,-0.0123) (2024.8750,-0.0077) (2024.9583,-0.0300) (2025.0417,-0.0035) (2025.1250,-0.0175) (2025.2083,0.0194) (2025.2917,-0.0351) (2025.3750,-0.0155) (2025.4583,-0.0624) (2025.5417,-0.0232) (2025.6250,-0.0337) (2025.7083,0.0376) (2025.7917,0.0313) (2025.8750,0.0325) (2025.9583,-0.0065) (2026.0417,-0.0053) (2026.1250,0.0124) (2026.2083,0.0314) (2026.2917,0.0431) (2026.3750,0.0488)};
\addlegendentry{GPT-5.4-mini (frontier)}
\addplot[gray!60!black, line width=0.7pt] coordinates {(2018.0417,-0.0270) (2018.1250,-0.0490) (2018.2083,-0.0152) (2018.2917,-0.0422) (2018.3750,-0.0409) (2018.4583,-0.0525) (2018.5417,-0.0016) (2018.6250,0.0607) (2018.7083,0.0416) (2018.7917,0.0475) (2018.8750,0.0651) (2018.9583,0.1114) (2019.0417,0.1170) (2019.1250,0.0545) (2019.2083,0.0380) (2019.2917,0.0085) (2019.3750,0.0094) (2019.4583,0.0369) (2019.5417,0.0227) (2019.6250,0.0044) (2019.7083,-0.0827) (2019.7917,-0.0972) (2019.8750,-0.0449) (2019.9583,-0.0279) (2020.0417,0.0243) (2020.1250,0.0019) (2020.2083,0.0085) (2020.2917,-0.0134) (2020.3750,0.0386) (2020.4583,0.0906) (2020.5417,0.1099) (2020.6250,0.1031) (2020.7083,0.0223) (2020.7917,-0.0195) (2020.8750,-0.0426) (2020.9583,-0.0146) (2021.0417,0.0233) (2021.1250,-0.0106) (2021.2083,0.0251) (2021.2917,0.0370) (2021.3750,0.0629) (2021.4583,0.0704) (2021.5417,0.0269) (2021.6250,0.0060) (2021.7083,-0.0115) (2021.7917,-0.0370) (2021.8750,-0.0263) (2021.9583,-0.0363) (2022.0417,0.0137) (2022.1250,0.0156) (2022.2083,0.0253) (2022.2917,-0.0031) (2022.3750,-0.0190) (2022.4583,-0.0290) (2022.5417,-0.0505) (2022.6250,-0.0781) (2022.7083,-0.0930) (2022.7917,-0.0718) (2022.8750,-0.0173) (2022.9583,-0.0263) (2023.0417,-0.0011) (2023.1250,0.0001) (2023.2083,0.0077) (2023.2917,-0.0254) (2023.3750,-0.0269) (2023.4583,0.0419) (2023.5417,0.0818) (2023.6250,0.0468) (2023.7083,-0.0005) (2023.7917,-0.0246) (2023.8750,-0.0124) (2023.9583,0.0421) (2024.0417,0.0411) (2024.1250,0.0269) (2024.2083,-0.0162) (2024.2917,-0.0094) (2024.3750,0.0290) (2024.4583,0.0608) (2024.5417,0.0355) (2024.6250,0.0278) (2024.7083,-0.0380) (2024.7917,-0.0390) (2024.8750,-0.0555) (2024.9583,-0.0554) (2025.0417,-0.0246) (2025.1250,-0.0171) (2025.2083,-0.0122) (2025.2917,-0.0229) (2025.3750,-0.0352) (2025.4583,-0.0457) (2025.5417,-0.0638) (2025.6250,-0.0549) (2025.7083,-0.0222) (2025.7917,0.0034) (2025.8750,-0.0268) (2025.9583,-0.0132) (2026.0417,-0.0050) (2026.1250,0.0240) (2026.2083,0.0099) (2026.2917,0.0095) (2026.3750,0.0190)};
\addlegendentry{FinBERT (clean)}
\addplot[gray!60!black, line width=0.7pt, densely dotted] coordinates {(2018.0417,0.1956) (2018.1250,0.1194) (2018.2083,0.1200) (2018.2917,0.1020) (2018.3750,0.0873) (2018.4583,0.0682) (2018.5417,0.0004) (2018.6250,0.0406) (2018.7083,0.0134) (2018.7917,0.0260) (2018.8750,0.0273) (2018.9583,0.0124) (2019.0417,0.0335) (2019.1250,-0.0238) (2019.2083,0.0194) (2019.2917,0.0066) (2019.3750,-0.0132) (2019.4583,-0.0239) (2019.5417,-0.0686) (2019.6250,0.0076) (2019.7083,-0.0159) (2019.7917,-0.0078) (2019.8750,-0.0001) (2019.9583,-0.0171) (2020.0417,0.0338) (2020.1250,0.0419) (2020.2083,0.0167) (2020.2917,0.0099) (2020.3750,-0.0133) (2020.4583,0.0129) (2020.5417,-0.0097) (2020.6250,0.0349) (2020.7083,0.0000) (2020.7917,0.0363) (2020.8750,-0.0242) (2020.9583,-0.0000) (2021.0417,-0.0138) (2021.1250,-0.0009) (2021.2083,0.0460) (2021.2917,0.0187) (2021.3750,0.0323) (2021.4583,0.0148) (2021.5417,0.0326) (2021.6250,0.0195) (2021.7083,0.0131) (2021.7917,-0.0126) (2021.8750,-0.0207) (2021.9583,-0.0046) (2022.0417,0.0380) (2022.1250,0.0265) (2022.2083,0.0123) (2022.2917,-0.0156) (2022.3750,-0.0196) (2022.4583,-0.0066) (2022.5417,-0.0000) (2022.6250,0.0126) (2022.7083,0.0038) (2022.7917,-0.0080) (2022.8750,0.0349) (2022.9583,0.0021) (2023.0417,0.0261) (2023.1250,0.0022) (2023.2083,0.0405) (2023.2917,-0.0232) (2023.3750,-0.0398) (2023.4583,-0.0149) (2023.5417,0.0384) (2023.6250,0.0223) (2023.7083,-0.0323) (2023.7917,-0.0394) (2023.8750,-0.0247) (2023.9583,0.0431) (2024.0417,0.0267) (2024.1250,0.0384) (2024.2083,0.0084) (2024.2917,0.0397) (2024.3750,0.0327) (2024.4583,0.0469) (2024.5417,0.0333) (2024.6250,0.0292) (2024.7083,0.0017) (2024.7917,0.0022) (2024.8750,0.0048) (2024.9583,-0.0040) (2025.0417,0.0085) (2025.1250,-0.0100) (2025.2083,-0.0013) (2025.2917,-0.0275) (2025.3750,-0.0275) (2025.4583,-0.0505) (2025.5417,-0.0562) (2025.6250,-0.0327) (2025.7083,0.0020) (2025.7917,0.0072) (2025.8750,-0.0272) (2025.9583,-0.0543) (2026.0417,-0.0421) (2026.1250,-0.0082) (2026.2083,0.0266) (2026.2917,0.0354) (2026.3750,0.0495)};
\addlegendentry{GPT-2 fine-tuned (clean)}
\end{axis}
\end{tikzpicture}
\caption{Monthly rank IC of each rater's rating of the same GPT-5.4-mini daily summaries with the realized 1-month excess return, by formation month (three-month centred moving average), with the two frontier raters' training cutoffs marked. A formation month counts as post-cutoff when its target window ends after the cutoff, so Table~\ref{tab:leak}'s post-cutoff months begin one horizon before each line.}
\Description{Line chart of monthly rank IC for up to four text raters on identical summaries, 2018 to 2026, with vertical dashed lines at the Claude Sonnet 4 and GPT-5.4-mini training cutoffs.}
\label{fig:cutoff}
\end{figure}
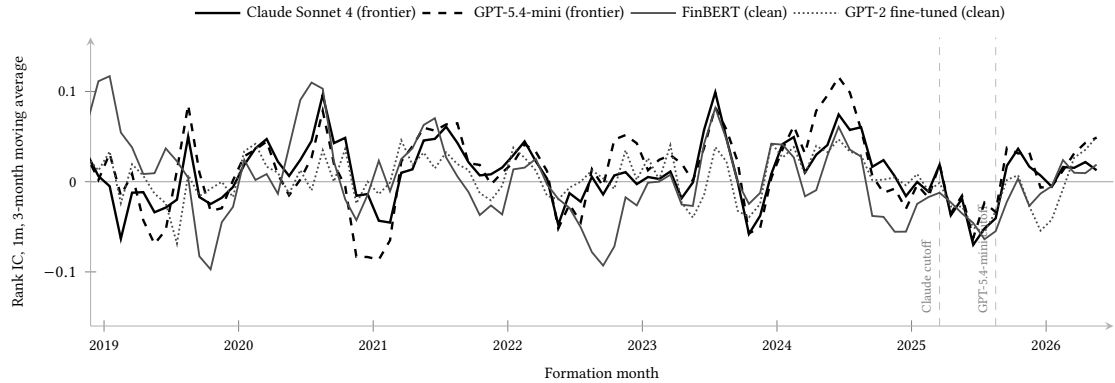

Table~\ref{tab:leak} lists every rater pair, cutoff and horizon that was run except the frontier--frontier pair at three and twelve months, which its note reports. Ratings are the carried-forward value at each month end within a company's observed span. The split is target-end-after-cutoff: a formation month is post-cutoff when its $h\times21$-session target window ends after the cutoff, which places the post windows' formation dates up to $h$ months before the cutoff, so the post-window returns partly precede it and, at twelve months, only two windows lie wholly after Claude's cutoff and none after GPT-5.4-mini's; it is the convention fixed when the test was designed, chosen so that post-cutoff windows exist at every horizon. The alternative split, formation after cutoff, leaves nine to fourteen months at one month and at most two at twelve, which is why the test as run tests leakage at one and three months and cannot establish its absence. Figure~\ref{fig:cutoff} smooths the monthly series with a centred three-month average for legibility; every statistic uses the unsmoothed months. Table~\ref{tab:prompt} is the prompted-forecasting table of Section~\ref{sec:leak}. Table~\ref{tab:prov} records what is documented about each model and what is not; where a snapshot, call date or setting was not logged it is marked unavailable rather than inferred. Table~\ref{tab:panels} gives the calendar bounds of the two panels; the rating panel begins when the daily summaries begin (January 2018) and the forecasting panel is the 2020--2025 evaluation window on a fixed-seed sample of 300 companies, which is why their month counts differ.

\begin{table}[!htbp]
\centering
\caption{Model provenance by role. ``Cutoff'' is the vendor's documented training-data or knowledge date; ``run'' is when this study invoked the model. Unavailable means not logged by the pipeline and not recoverable.}
\label{tab:prov}
{\scriptsize
\begin{tabular}{@{}p{1.9cm}p{2.9cm}p{2.3cm}p{2.4cm}p{3.8cm}@{}}
\toprule
Model (role) & Release; snapshot & Documented cutoff & Run in this study & What the row establishes \\
\midrule
GPT-2 small (clean rater; text-tower representation) & 14 Feb 2019; \texttt{gpt2}, 124M & WebText, which excludes links created after Dec 2017 & Fine-tuned 1--2 Oct 2026; scoring 2--3 Oct; embeddings of the news-and-social panel 2--3 Oct, of the all-source panel 4--5 Oct 2026 & Pretraining precedes the 2018--2026 rating panel. The fine-tuning records (2010--2019, rule-labelled from their own text, not from outcomes) precede the 2020--2025 ranking evaluation but overlap the first two years of the rating panel, so 24 formation months of the rating panel overlap the fine-tuning period; those comparisons have training-period overlap, while the 2020--2025 evaluation is cleanly separated. \\
FinBERT (second clean rater; production polarity columns) & Checkpoint \texttt{ProsusAI/finbert}, loaded by name; weights as described by \citet{araci2019finbert} & BERT-base further pretrained on a financial subset of TRC2 (Reuters; January 2008 to February 2009 per the distributor, \citealt{nist2009trc2}, which Araci describes as 2008--2010) and fine-tuned on the Financial PhraseBank (2014), per \citet{araci2019finbert}; that the loaded checkpoint is those weights is not independently verifiable & June 2026 (production pipeline) & ``Clean'' rests on the published description, not on a verified training date of the checkpoint; the GPT-2 rater is the clean rater whose data are fully documented. \\
Claude Sonnet 4 (rater of summaries, P6; narrative generator) & 22 May 2025; \texttt{claude-sonnet-4-}\allowbreak\texttt{20250514} & Training data through March 2025 (the vendor's ``reliable knowledge cutoff'' is January 2025); the later, conservative date 31 March 2025 is used & Rating 25 June 2026 through the Databricks Foundation Model API endpoint \texttt{databricks-}\allowbreak\texttt{claude-}\allowbreak\texttt{sonnet-4}, a partner platform with its own retirement schedule, temperature 0, per the production run record; narrative generation June 2026, temperature 0, exact call dates not logged (prompt archived 27 Sept 2026) & Rater-clean comparisons only: it rates summaries written by a 2026 model. \\
GPT-5.4-mini (summarizer, P3; rater, P6; prompted forecaster) & 17 March 2026; \texttt{gpt-5.4-mini-}\allowbreak\texttt{2026-03-17} & 31 Aug 2025 (documented knowledge cutoff) & Summaries June 2026, call dates unavailable, temperature 0.1; re-rating 2 Oct 2026, temperature 0; forecasting 3--4 Oct 2026 & The comparison with a common summarizer--rater cutoff (summarizer is the rater); its post windows are nine to ten months. \\
Qwen3-Embedding-0.6B (text-tower representation) & 5 June 2025 & Not disclosed & Embeddings 4 Oct 2026 & Nothing about leakage; its tower is judged against the within-company identity placebo, not a cutoff. \\
GPT-5 (judge of narratives) & 7 Aug 2025; snapshot unavailable & 30 Sept 2024 (knowledge cutoff per the API model documentation, \citealt{openai2025gpt5docs}; not used, faithfulness task) & Temperature 0 (from the prompt record); snapshot, call date and reasoning settings unavailable; verdict files dated 27 Sept 2026 & Agreement with the deterministic rules only (Table~\ref{tab:faith}). \\
\bottomrule
\end{tabular}}
\end{table}

\begin{table}[!htbp]
\centering
\caption{Calendar bounds, month counts and baseline rankings of every evaluation sample in the paper. Formation months are month ends; a month enters when at least twenty names are rated or scored and its target has resolved by June 2026. The last column is the hybrid's neutral IC on that sample at that horizon (market-only in parentheses), so every increment in the paper can be read against its own baseline. ``Wholly after'' counts the post months whose formation date itself follows the cutoff.}
\label{tab:panels}
{\scriptsize
\resizebox{\linewidth}{!}{\begin{tabular}{@{}lllrrrl@{}}
\toprule
Sample & $h$ & Formation months & Months & Post, Claude (wholly after) & Post, GPT-5.4-mini (wholly after) & Hybrid (market-only) neutral IC \\
\midrule
Ledger, full evaluation panel (Table~\ref{tab:ledger}) & 1m & Jan 2020 -- Dec 2025 & 72 & -- & -- & 0.028 (0.020) \\
 & 3m & Jan 2020 -- Dec 2025 & 72 & -- & -- & 0.044 (0.027) \\
 & 12m & Jan 2020 -- May 2025 & 65 & -- & -- & 0.093 (0.063) \\
Text-covered rows, all sources (Table~\ref{tab:texttower}; 41.5--43.7\% of rows) & 12m & Jan 2020 -- May 2025 & 65 & -- & -- & 0.110 (0.057) \\
Scientific-text rows (Table~\ref{tab:lenses}, L6b--c; 33\%) & 3m & Jan 2020 -- Dec 2025 & 72 & -- & -- & 0.046 \\
 & 12m & Jan 2020 -- May 2025 & 65 & -- & -- & 0.105 \\
Rating (Table~\ref{tab:leak}; all rated companies) & 1m & Jan 2018 -- May 2026 & 101 & 15 (14) & 10 (9) & -- \\
 & 3m & Jan 2018 -- Feb 2026 & 98 & 15 (11) & 9 (6) & -- \\
 & 12m & Jan 2018 -- May 2025 & 89 & 15 (2) & 10 (0) & -- \\
Forecasting (Table~\ref{tab:prompt}; 300 companies, 15{,}627 rows) & 1m & Jan 2020 -- Dec 2025 & 72 & -- & 5 (4) & 0.059 (0.060) \\
 & 3m & Jan 2020 -- Dec 2025 & 72 & -- & 7 (4) & 0.087 (0.067) \\
 & 12m & Jan 2020 -- May 2025 & 65 & -- & 10 (0) & 0.183 (0.096) \\
\bottomrule
\end{tabular}}}
\end{table}

\begin{table}[!htbp]
\centering
\caption{Identity-blind prompted forecasting, GPT-5.4-mini, 300 sampled companies, 2020--2025 (65, 72 and 72 months): monthly rank IC of the blind and revealed scores against the realized excess return, raw and price- and liquidity-neutral (Newey--West $t$), beside the market score and the hybrid prediction on the same rows; the within-month rank correlation of the blind score with the hybrid; the revealed-minus-blind gap before and after the model's cutoff (31 August 2025; post months have target windows ending after it) and the before-minus-after difference; and the top- and bottom-decile hit rates against XBI for the blind score and the hybrid.}
\label{tab:prompt}
{\footnotesize\setlength{\tabcolsep}{3pt}
\begin{tabular}{lrrrrrrrrr}
\toprule
 & \multicolumn{2}{c}{Blind} & \multicolumn{2}{c}{Revealed} & Market / hybrid & Overlap & \multicolumn{2}{c}{Revealed $-$ blind, raw} & D1 / D10 hit rate \\
\cmidrule(lr){2-3}\cmidrule(lr){4-5}\cmidrule(lr){8-9}
$h$ & Raw ($t$) & Neutral ($t$) & Raw & Neutral & neutral IC & blind$\sim$hybrid & Pre / post & Pre $-$ post ($t$) & blind; hybrid \\
\midrule
12m & $0.134$ (3.6) & $0.078$ (3.3) & $0.155$ & $0.104$ & 0.096 / 0.183 & 0.61 & $+0.025$ / $+0.008$ & $+0.017$ (1.7) & 47 / 33\%; 51 / 27\% \\
3m & $0.081$ (3.4) & $0.050$ (3.0) & $0.093$ & $0.061$ & 0.067 / 0.087 & 0.59 & $+0.013$ / $+0.001$ & $+0.012$ (2.5) & 46 / 40\%; 48 / 35\% \\
1m & $0.062$ (3.3) & $0.039$ (2.6) & $0.069$ & $0.044$ & 0.060 / 0.059 & 0.61 & $+0.007$ / $-0.001$ & $+0.008$ (0.8) & 47 / 41\%; 49 / 39\% \\
\bottomrule
\end{tabular}}
\par\medskip\begin{minipage}{\linewidth}\footnotesize\raggedright\setlength{\parskip}{1pt}
\par 99.9\% of calls returned a parsable integer (92 distinct values). Pre / post months: 55 / 10, 65 / 7, 67 / 5 at twelve, three and one months. The blind score correlates 0.63 with the price-model percentile line; its leave-one-year-out blend weight into the hybrid is 0 at twelve and one months (0 or 0.1 at three); its deciles match the hybrid's exactly for 20\% of company-months (10\% by chance). Snapshot \texttt{gpt-5.4-mini-2026-03-17}, temperature 0, low reasoning effort, 3--4 October 2026.
\end{minipage}
\end{table}

%%FIG-CUTOFF-BEGIN
%%FIG-CUTOFF-END

\FloatBarrier
\section{Faithfulness Rules and Judge Disagreements}
\label{app:faith}

The checker parses each narrative's JSON (fields \texttt{horizon\_summary}, \texttt{analog\_context}, \texttt{cohort\_confidence}) and the exact input payload the generator received, and applies the six rules of Section~\ref{sec:leak}. Operational details: a numeric token matches an input exactly, after rounding the input to one to six decimals, or as the length of an input list or cohort, and a value equal to an input times 100 is counted separately as ``converted'' because the rulebook forbids arithmetic; the entity rule excludes sentence-initial common words and a stop-list of schema words (Market, Sentiment, Scientific, Fundamental, Macro, IQR, \ldots); the forecast lexicon is will, would, expect(s/ed), likely, unlikely, should, forecast, predict, projected, anticipate, poised, set to, upside, downside, outperform, underperform, buy, sell, bullish, bearish, target price, recommend, catalyst ahead, going forward, with a negated mention recorded and not counted; partially null variables are judged field by field.

Results on 300 narratives: 6{,}252 numeric tokens, all exact matches; 0 untraced entities; 0 forecast hits (3 negated mentions); 1{,}599 null fields, none given a value; 0 evaluative hits; 79 narratives (28, 25, 26 by horizon) fail rule (3) on the single word ``regulatory'' in ``scientific and regulatory event data is unavailable''. One three-month narrative (TKNO) failed the generator's schema validation in the generation run and is recovered field by field here. The GPT-5 judge, run on 299 narratives with the generator's exact input and the two defining rules, flagged 14 of the 79 and 0 of the 220 rule-passing narratives it judged (one unjudged); by horizon 2 of 28, 5 of 25, 7 of 26. The 65 misses and 14 flags, with excerpts, are available with the code. A supplementary scan for could, may, on track, outlook and risk of, forbidden by the prompt but absent from rule (4)'s lexicon, finds 0 occurrences in the 300 narratives.

\section{Text Tower and Regime Blend Tables}
\label{app:tables}
\paragraph{Text-tower implementation, as run.} Each company-month's text (records in the trailing 30 days, most recent first, source-tagged, cut at 6{,}000 characters) is embedded once by each frozen model. Pre-2020 model: Hugging Face \texttt{gpt2} (124M) with its byte-level BPE tokenizer, pad token set to the end-of-text token, inputs truncated to 512 tokens (so a long month keeps its most recent records), and the vector is the mean of the last hidden state over non-pad tokens, 768 dimensions, stored in float16 without normalization. Current model: the \texttt{databricks-qwen3-embedding-0-6b} serving endpoint (Qwen3-Embedding-0.6B), input the same text cut at 6{,}000 characters with no instruction prefix, 1{,}024 dimensions stored in float16; pooling and any normalization are the endpoint's own and were not controlled client-side. Head: for test year $Y$ the training rows are years through $Y-2$ (year $Y-1$ is excluded entirely, so no twelve-month target overlaps the test year); features are standardized on the training rows; a ridge regression on the within-month demeaned excess return is fitted with the penalty chosen from $\{10, 100, 1000, 10000\}$ by five-fold group cross-validation on the training rows with months as groups, maximizing the mean monthly Spearman IC of the out-of-fold predictions; the chosen model is refitted on all training rows and its predictions for year $Y$ are ranked within month. The two placebos permute the embedding rows, within month across companies or within company across months, before the same head is fitted.

Table~\ref{tab:texttower} is the two-generation text-tower table of Section~\ref{sec:clean}; Table~\ref{tab:regime} the leave-one-year-out regime blend of Section~\ref{sec:state}; Table~\ref{tab:years} the year-by-year gains; Table~\ref{tab:threeway} the three-way regime blend of H8; Table~\ref{tab:states} the state-month composition and permutation details; Table~\ref{tab:stategrid} every channel on the state grid.

\begin{table}[!htbp]
\centering
\caption{Text-only rankings from a 2019 and a 2025 model on the same source records and downstream protocol, evaluation rows with text, 2020--2025 (72, 72 and 65 months): monthly rank IC alone, raw and price- and liquidity-neutral (Newey--West $t$; neutral block-bootstrap interval); the within-company shuffle placebo (neutral); the within-month rank correlation with the hybrid; and the change in neutral IC from adding the text ranking to the market-only ranking and to the hybrid, as a plain average (weight 0.5) and with the weight chosen leave-one-year-out on the other five years (paired, $t$; for the blend with the hybrid also the paired block-bootstrap 95\% interval). Panel A gives the standalone rankings and both placebos, Panel B the blends with the market-only ranking, Panel C the blends with the hybrid. The standalone intervals in Panel A are not intervals on the increments.}
\label{tab:texttower}
{\footnotesize\setlength{\tabcolsep}{4pt}
\textbf{A. Standalone text rankings and placebos (neutral unless marked)}\par\smallskip
\begin{tabular}{@{}llrrrrr@{}}
\toprule
$h$ & Model & Raw ($t$) & Neutral ($t$; boot 95\%) & Identity placebo ($t$) & Within-month placebo ($t$) & Overlap w/ hybrid \\
\midrule
12m & GPT-2 (2019) & $0.120$ (6.7) & $+0.039$ (4.1; $[+0.019, +0.055]$) & $+0.036$ (4.1) & $-0.010$ ($-2.6$) & 0.33 \\
12m & Qwen3-Embedding (2025) & $0.113$ (6.5) & $+0.034$ (3.4; $[+0.014, +0.052]$) & $+0.041$ (3.8) & $+0.003$ (0.3) & 0.34 \\
3m & GPT-2 (2019) & $0.077$ (6.5) & $+0.030$ (4.4; $[+0.020, +0.039]$) & $+0.042$ (6.2) & $-0.003$ ($-0.5$) & 0.26 \\
3m & Qwen3-Embedding (2025) & $0.072$ (6.7) & $+0.019$ (2.7; $[+0.005, +0.036]$) & $+0.029$ (4.0) & $+0.006$ (1.2) & 0.30 \\
1m & GPT-2 (2019) & $0.030$ (3.9) & $+0.005$ (0.8; $[-0.007, +0.019]$) & $+0.014$ (2.5) & $-0.001$ ($-0.2$) & 0.15 \\
1m & Qwen3-Embedding (2025) & $0.029$ (3.6) & $+0.000$ (0.0; $[-0.010, +0.009]$) & $+0.014$ (2.1) & $-0.004$ ($-0.7$) & 0.18 \\
\bottomrule
\end{tabular}
\par\medskip
\textbf{B. Text ranking added to the market-only ranking: change in neutral IC ($t$)}\par\smallskip
\begin{tabular}{@{}llrr@{}}
\toprule
$h$ & Model & Plain average (weight 0.5) & Leave-one-year-out weight ($w$ range) \\
\midrule
12m & GPT-2 (2019) & $-0.000$ ($-0.0$) & $+0.001$ (0.2; 0.1--0.3) \\
12m & Qwen3-Embedding (2025) & $-0.006$ ($-0.5$) & $+0.003$ (0.7; 0.1--0.2) \\
3m & GPT-2 (2019) & $+0.013$ (1.7) & $+0.007$ (1.0; 0.2--0.6) \\
3m & Qwen3-Embedding (2025) & $+0.001$ (0.1) & $+0.000$ (0.0; 0.1--0.4) \\
1m & GPT-2 (2019) & $-0.007$ ($-1.1$) & $+0.002$ (1.2; 0.1) \\
1m & Qwen3-Embedding (2025) & $-0.009$ ($-1.3$) & $-0.001$ ($-1.6$; 0--0.1) \\
\bottomrule
\end{tabular}
\par\medskip
\textbf{C. Text ranking added to the hybrid: change in neutral and raw IC ($t$)}\par\smallskip
\begin{tabular}{@{}llrrr@{}}
\toprule
$h$ & Model & Plain average, $\Delta$neutral & LOYO, $\Delta$neutral ($t$; boot 95\%; $w$) & LOYO, $\Delta$raw ($t$) \\
\midrule
12m & GPT-2 (2019) & $-0.035$ ($-3.0$) & $+0.003$ (1.5; $[-0.001, +0.007]$; 0.1) & $+0.001$ (1.3) \\
12m & Qwen3-Embedding (2025) & $-0.040$ ($-3.4$) & $+0.003$ (1.0; $[-0.002, +0.008]$; 0.1) & $+0.001$ (0.7) \\
3m & GPT-2 (2019) & $-0.002$ ($-0.3$) & $+0.004$ (2.0; $[+0.002, +0.005]$; 0.1) & $+0.002$ (3.1) \\
3m & Qwen3-Embedding (2025) & $-0.011$ ($-1.5$) & $+0.003$ (1.5; $[-0.001, +0.006]$; 0.1) & $+0.001$ (1.7) \\
1m & GPT-2 (2019) & $-0.015$ ($-2.4$) & $-0.002$ ($-1.7$; $[-0.004, +0.000]$; 0--0.1) & $-0.001$ ($-1.1$) \\
1m & Qwen3-Embedding (2025) & $-0.018$ ($-2.7$) & $0$ (weight 0 in every fold) & $0$ \\
\bottomrule
\end{tabular}}
\par\medskip\begin{minipage}{\linewidth}\footnotesize\raggedright\setlength{\parskip}{1pt}
\par Text coverage of evaluation rows 41.5\% at twelve months and 43.7\% at one and three. On the text rows the market-only ranking's neutral IC is 0.019 / 0.022 / 0.057 and the hybrid's 0.031 / 0.043 / 0.110 at one, three and twelve months (raw 0.120 / 0.168 / 0.244 and 0.122 / 0.166 / 0.265). Plain averaging also lowers raw IC in every cell ($-0.020$ to $-0.038$ against the hybrid). Weights for the leave-one-year-out blend are chosen on neutral IC; choosing them on raw IC gives the same cells within 0.003. The 2019 model's rows use its own 768-dimensional representation of the same text; its earlier run on news and social text alone (45{,}670 company-months) gave $+0.029$ neutral at twelve months with the identity placebo at $+0.042$.
\end{minipage}
\end{table}

\begin{table}[!htbp]
\centering
\caption{Leave-one-year-out regime blend, twelve-month horizon, 65 months: neutral and raw rank IC of each ranking, and the paired regime-minus-static and regime-minus-full-model differences with Newey--West $t$ (lags 12) and block-bootstrap 95\% interval. Weights per state tercile are fitted on the other five years. The chosen weight on fundamentals is 0 in the low tercile and 1 in the high tercile of trailing return and of realized volatility in all six folds.}
\label{tab:regime}
{\footnotesize
\resizebox{\linewidth}{!}{\begin{tabular}{lrrrrrr}
\toprule
Ranking & Neutral IC & Raw IC & vs static, neutral ($t$; boot 95\%) & vs static, raw ($t$) & vs full model, neutral ($t$) & months $>$ static \\
\midrule
Market-only & 0.063 & 0.289 & & & & \\
Static LOYO blend of the two channels & 0.084 & 0.295 & & & & \\
Full hybrid model (all families) & 0.093 & 0.298 & & & & \\
Regime blend, state $=$ trailing 12m XBI return & 0.097 & 0.299 & $+0.013$ (2.4; $[+0.003, +0.024]$) & $+0.004$ (1.8) & $+0.004$ (0.8) & 37/65 \\
Regime blend, state $=$ realized volatility & 0.097 & 0.300 & $+0.013$ (2.2; $[+0.002, +0.024]$) & $+0.005$ (1.9) & $+0.004$ (0.6) & 43/65 \\
Regime blend, joint state & 0.099 & 0.300 & $+0.015$ (2.7; $[+0.004, +0.026]$) & $+0.005$ (2.0) & $+0.006$ (0.9) & 46/65 \\
Regime blend, state $=$ high-yield spread & 0.079 & 0.293 & $-0.005$ ($-1.8$) & $-0.003$ ($-2.3$) & $-0.014$ ($-2.3$) & 18/65 \\
\bottomrule
\end{tabular}}}
\par\medskip\begin{minipage}{\linewidth}\footnotesize\raggedright\setlength{\parskip}{1pt}
\par At three months the trailing-return blend gains $+0.007$ over static ($t$ 2.4) and the full model still exceeds every blend; at one month the fundamental channel is never admitted. The static weight wanders across folds (0.2--0.7), so the regime-minus-full-model line is the stricter comparison.
\end{minipage}
\end{table}

\begin{table}[!htbp]
\centering
\caption{Twelve-month neutral gain over the market-only ranking by evaluation year: full model, fundamental family model and sentiment-pipeline family model (whose admitted columns are the attention counts), mean of the monthly paired differences within each year.}
\label{tab:years}
{\footnotesize
\begin{tabular}{@{}lrrrr@{}}
\toprule
Year & Months & Full model & Fundamentals & Attention \\
\midrule
2020 & 12 & $+0.047$ & $+0.048$ & $0.000$ (no admitted column) \\
2021 & 12 & $+0.014$ & $+0.015$ & $+0.012$ \\
2022 & 12 & $+0.048$ & $+0.016$ & $+0.051$ \\
2023 & 12 & $+0.043$ & $+0.032$ & $+0.029$ \\
2024 & 12 & $+0.024$ & $+0.026$ & $+0.010$ \\
2025 (to May) & 5 & $-0.028$ & $-0.033$ & $+0.032$ \\
\bottomrule
\end{tabular}}
\end{table}

\FloatBarrier
{\footnotesize\setlength{\tabcolsep}{4pt}
\begin{longtable}{@{}lrrrrr@{}}
\caption{The three-way regime blend (market-only, hybrid and a language-model text ranking; weights per state tercile chosen leave-one-year-out on the other five years, 66 triples) against the hybrid: neutral IC of the blend, paired neutral difference with Newey--West $t$ and block-bootstrap 95\% interval, months with a positive difference, and the paired raw difference. H8 asks whether any twelve-month row beats the hybrid.}\label{tab:threeway} \\
\toprule
State & Blend neutral IC & vs hybrid, neutral ($t$) & boot 95\% & months $>0$ & vs hybrid, raw ($t$) \\
\midrule
\endfirsthead
\multicolumn{6}{@{}l}{\emph{Table~\thetable, continued}} \\
\toprule
State & Blend neutral IC & vs hybrid, neutral ($t$) & boot 95\% & months $>0$ & vs hybrid, raw ($t$) \\
\midrule
\endhead
\midrule
\multicolumn{6}{r@{}}{\emph{continued on the next page}} \\
\endfoot
\bottomrule
\endlastfoot
\multicolumn{6}{@{}l}{\emph{All text, current model, 12m, all rows, hybrid rank where no text}} \\
static (no state) & 0.086 & $-0.007$ (-1.5) & $[-0.015, +0.002]$ & 36/65 & $-0.034$ (-3.4) \\
trailing return & 0.089 & $-0.004$ (-1.2) & $[-0.011, +0.003]$ & 29/65 & $-0.020$ (-1.9) \\
realized volatility & 0.093 & $-0.000$ (-0.0) & $[-0.004, +0.005]$ & 30/65 & $-0.021$ (-2.8) \\
joint & 0.084 & $-0.009$ (-2.1) & $[-0.017, -0.001]$ & 33/65 & $-0.029$ (-3.2) \\
high-yield spread & 0.095 & $+0.002$ (0.4) & $[-0.008, +0.014]$ & 37/65 & $-0.034$ (-3.9) \\
\multicolumn{6}{@{}l}{\emph{All text, current model, 12m, rows with text}} \\
static (no state) & 0.112 & $+0.003$ (1.0) & $[-0.002, +0.008]$ & 36/65 & $+0.001$ (0.7) \\
trailing return & 0.111 & $+0.001$ (0.4) & $[-0.004, +0.008]$ & 27/65 & $+0.001$ (0.9) \\
realized volatility & 0.107 & $-0.003$ (-1.2) & $[-0.009, +0.002]$ & 21/65 & $-0.002$ (-1.1) \\
joint & 0.108 & $-0.002$ (-0.4) & $[-0.011, +0.007]$ & 32/65 & $+0.001$ (0.5) \\
high-yield spread & 0.104 & $-0.006$ (-1.0) & $[-0.019, +0.005]$ & 25/65 & $-0.004$ (-1.4) \\
\multicolumn{6}{@{}l}{\emph{All text, current model, 3m, all rows, hybrid rank where no text}} \\
static (no state) & 0.044 & $+0.000$ (0.0) & $[-0.011, +0.011]$ & 34/72 & $-0.030$ (-3.4) \\
trailing return & 0.052 & $+0.008$ (1.6) & $[-0.002, +0.020]$ & 36/72 & $-0.029$ (-3.5) \\
realized volatility & 0.041 & $-0.002$ (-0.5) & $[-0.011, +0.005]$ & 30/72 & $-0.029$ (-3.3) \\
joint & 0.049 & $+0.005$ (0.9) & $[-0.007, +0.017]$ & 32/72 & $-0.029$ (-3.9) \\
high-yield spread & 0.047 & $+0.003$ (0.6) & $[-0.010, +0.016]$ & 35/72 & $-0.030$ (-3.5) \\
\multicolumn{6}{@{}l}{\emph{All text, current model, 3m, rows with text}} \\
static (no state) & 0.046 & $+0.003$ (1.5) & $[-0.001, +0.006]$ & 41/72 & $+0.001$ (1.7) \\
trailing return & 0.038 & $-0.005$ (-2.1) & $[-0.011, -0.000]$ & 35/72 & $-0.009$ (-1.2) \\
realized volatility & 0.042 & $-0.002$ (-0.7) & $[-0.006, +0.002]$ & 39/72 & $+0.001$ (0.4) \\
joint & 0.041 & $-0.002$ (-0.7) & $[-0.007, +0.003]$ & 37/72 & $-0.001$ (-0.3) \\
high-yield spread & 0.029 & $-0.015$ (-2.2) & $[-0.028, -0.003]$ & 26/72 & $-0.010$ (-1.1) \\
\multicolumn{6}{@{}l}{\emph{All text, pre-2020 model, 12m, all rows, hybrid rank where no text}} \\
static (no state) & 0.092 & $-0.001$ (-0.3) & $[-0.007, +0.004]$ & 36/65 & $-0.013$ (-1.4) \\
trailing return & 0.088 & $-0.005$ (-1.8) & $[-0.010, +0.001]$ & 31/65 & $-0.019$ (-2.1) \\
realized volatility & 0.092 & $-0.001$ (-0.2) & $[-0.006, +0.004]$ & 37/65 & $-0.022$ (-2.5) \\
joint & 0.086 & $-0.007$ (-1.9) & $[-0.016, -0.000]$ & 31/65 & $-0.022$ (-2.8) \\
high-yield spread & 0.092 & $-0.001$ (-0.3) & $[-0.007, +0.005]$ & 34/65 & $-0.025$ (-2.7) \\
\multicolumn{6}{@{}l}{\emph{All text, pre-2020 model, 12m, rows with text}} \\
static (no state) & 0.113 & $+0.003$ (1.5) & $[-0.001, +0.007]$ & 32/65 & $+0.001$ (1.3) \\
trailing return & 0.105 & $-0.005$ (-1.5) & $[-0.011, +0.002]$ & 24/65 & $-0.000$ (-0.3) \\
realized volatility & 0.109 & $-0.000$ (-0.1) & $[-0.008, +0.007]$ & 24/65 & $-0.000$ (-0.3) \\
joint & 0.108 & $-0.002$ (-0.5) & $[-0.014, +0.006]$ & 31/65 & $-0.000$ (-0.2) \\
high-yield spread & 0.108 & $-0.001$ (-0.4) & $[-0.010, +0.006]$ & 18/65 & $-0.001$ (-0.7) \\
\multicolumn{6}{@{}l}{\emph{All text, pre-2020 model, 3m, all rows, hybrid rank where no text}} \\
static (no state) & 0.057 & $+0.013$ (1.9) & $[+0.001, +0.025]$ & 42/72 & $-0.040$ (-4.5) \\
trailing return & 0.055 & $+0.012$ (1.8) & $[+0.001, +0.024]$ & 33/72 & $-0.027$ (-3.4) \\
realized volatility & 0.057 & $+0.013$ (1.9) & $[+0.001, +0.025]$ & 42/72 & $-0.040$ (-4.5) \\
joint & 0.054 & $+0.010$ (1.6) & $[+0.001, +0.021]$ & 36/72 & $-0.028$ (-3.4) \\
high-yield spread & 0.052 & $+0.008$ (1.5) & $[-0.001, +0.017]$ & 40/72 & $-0.032$ (-3.7) \\
\multicolumn{6}{@{}l}{\emph{All text, pre-2020 model, 3m, rows with text}} \\
static (no state) & 0.046 & $+0.002$ (1.2) & $[-0.000, +0.005]$ & 42/72 & $+0.003$ (2.2) \\
trailing return & 0.042 & $-0.001$ (-0.2) & $[-0.011, +0.008]$ & 38/72 & $-0.011$ (-1.3) \\
realized volatility & 0.039 & $-0.005$ (-1.2) & $[-0.013, +0.002]$ & 40/72 & $+0.000$ (0.1) \\
joint & 0.041 & $-0.003$ (-0.5) & $[-0.011, +0.005]$ & 39/72 & $-0.004$ (-0.9) \\
high-yield spread & 0.036 & $-0.007$ (-1.0) & $[-0.018, +0.002]$ & 32/72 & $-0.007$ (-0.7) \\
\multicolumn{6}{@{}l}{\emph{Scientific text, current model, 12m, all rows, hybrid rank where no text}} \\
static (no state) & 0.092 & $-0.001$ (-0.3) & $[-0.008, +0.006]$ & 31/65 & $-0.029$ (-3.7) \\
trailing return & 0.096 & $+0.002$ (0.8) & $[-0.003, +0.008]$ & 35/65 & $-0.020$ (-3.1) \\
realized volatility & 0.095 & $+0.002$ (0.5) & $[-0.006, +0.011]$ & 28/65 & $-0.026$ (-5.6) \\
joint & 0.090 & $-0.003$ (-0.8) & $[-0.009, +0.003]$ & 29/65 & $-0.028$ (-3.9) \\
high-yield spread & 0.104 & $+0.011$ (1.7) & $[+0.001, +0.024]$ & 42/65 & $-0.031$ (-3.7) \\
\multicolumn{6}{@{}l}{\emph{Scientific text, current model, 12m, rows with text}} \\
static (no state) & 0.107 & $+0.002$ (0.8) & $[-0.002, +0.007]$ & 33/65 & $+0.000$ (0.2) \\
trailing return & 0.105 & $+0.001$ (0.2) & $[-0.009, +0.009]$ & 29/65 & $+0.001$ (0.6) \\
realized volatility & 0.100 & $-0.004$ (-0.4) & $[-0.027, +0.013]$ & 28/65 & $-0.002$ (-0.4) \\
joint & 0.103 & $-0.002$ (-0.3) & $[-0.017, +0.010]$ & 30/65 & $+0.000$ (0.0) \\
high-yield spread & 0.102 & $-0.002$ (-0.3) & $[-0.016, +0.010]$ & 25/65 & $-0.001$ (-0.4) \\
\multicolumn{6}{@{}l}{\emph{Scientific text, current model, 3m, all rows, hybrid rank where no text}} \\
static (no state) & 0.061 & $+0.017$ (2.8) & $[+0.006, +0.030]$ & 49/72 & $-0.024$ (-4.5) \\
trailing return & 0.058 & $+0.014$ (2.3) & $[+0.003, +0.028]$ & 40/72 & $-0.017$ (-3.6) \\
realized volatility & 0.061 & $+0.017$ (2.8) & $[+0.006, +0.030]$ & 49/72 & $-0.024$ (-4.5) \\
joint & 0.060 & $+0.016$ (2.6) & $[+0.005, +0.029]$ & 47/72 & $-0.021$ (-4.5) \\
high-yield spread & 0.055 & $+0.012$ (2.1) & $[+0.004, +0.021]$ & 46/72 & $-0.017$ (-3.2) \\
\multicolumn{6}{@{}l}{\emph{Scientific text, current model, 3m, rows with text}} \\
static (no state) & 0.055 & $+0.009$ (2.3) & $[+0.005, +0.015]$ & 46/72 & $+0.004$ (2.5) \\
trailing return & 0.055 & $+0.009$ (1.7) & $[-0.002, +0.022]$ & 47/72 & $-0.006$ (-0.8) \\
realized volatility & 0.053 & $+0.007$ (1.9) & $[+0.002, +0.013]$ & 44/72 & $+0.004$ (1.6) \\
joint & 0.055 & $+0.009$ (2.5) & $[+0.003, +0.016]$ & 46/72 & $+0.001$ (0.4) \\
high-yield spread & 0.052 & $+0.006$ (1.2) & $[-0.002, +0.013]$ & 33/72 & $+0.002$ (1.0) \\
\end{longtable}}
\FloatBarrier

\begin{table}[!htbp]
\centering
\caption{State-month composition of the 65 twelve-month evaluation months (January 2020 to May 2025) by tercile of each pre-specified state, and the contiguous high and low episodes. Tercile cut points: trailing twelve-month XBI return $-7.9\%$ and $+18.4\%$; realized twelve-month volatility 24.6\% and 30.7\% annualized.}
\label{tab:states}
{\scriptsize
\begin{tabular}{@{}lrrrrrrrrrrrrl@{}}
\toprule
 & \multicolumn{6}{c}{Trailing return: low / mid / high} & \multicolumn{6}{c}{Realized volatility: low / mid / high} & \\
Year & \multicolumn{2}{c}{low} & \multicolumn{2}{c}{mid} & \multicolumn{2}{c}{high} & \multicolumn{2}{c}{low} & \multicolumn{2}{c}{mid} & \multicolumn{2}{c}{high} & Months \\
\midrule
2020 & \multicolumn{2}{c}{1} & \multicolumn{2}{c}{3} & \multicolumn{2}{c}{8} & \multicolumn{2}{c}{1} & \multicolumn{2}{c}{2} & \multicolumn{2}{c}{9} & 12 \\
2021 & \multicolumn{2}{c}{2} & \multicolumn{2}{c}{3} & \multicolumn{2}{c}{7} & \multicolumn{2}{c}{5} & \multicolumn{2}{c}{5} & \multicolumn{2}{c}{2} & 12 \\
2022 & \multicolumn{2}{c}{12} & \multicolumn{2}{c}{0} & \multicolumn{2}{c}{0} & \multicolumn{2}{c}{3} & \multicolumn{2}{c}{5} & \multicolumn{2}{c}{4} & 12 \\
2023 & \multicolumn{2}{c}{4} & \multicolumn{2}{c}{7} & \multicolumn{2}{c}{1} & \multicolumn{2}{c}{7} & \multicolumn{2}{c}{4} & \multicolumn{2}{c}{1} & 12 \\
2024 & \multicolumn{2}{c}{0} & \multicolumn{2}{c}{6} & \multicolumn{2}{c}{6} & \multicolumn{2}{c}{1} & \multicolumn{2}{c}{5} & \multicolumn{2}{c}{6} & 12 \\
2025 (to May) & \multicolumn{2}{c}{3} & \multicolumn{2}{c}{2} & \multicolumn{2}{c}{0} & \multicolumn{2}{c}{5} & \multicolumn{2}{c}{0} & \multicolumn{2}{c}{0} & 5 \\
\bottomrule
\end{tabular}}
\par\medskip\begin{minipage}{\linewidth}\footnotesize\raggedright\setlength{\parskip}{1pt}
\par Episodes (runs of three or more months in one tercile). Trailing return, high: May 2020 to June 2021 (14 months) and August to November 2024 (4); low: November 2021 to December 2022 (14) and September to November 2023 (3). Realized volatility, high: April 2020 to February 2021 (11), October to December 2022 (3) and April to August 2024 (5); low: April to June 2021 (3), November 2021 to March 2022 (5), April to October 2023 (7) and December 2024 to May 2025 (6). Permutation test (Section~\ref{sec:state}): the 65 state labels are rolled by a uniform random circular shift, cut into blocks of twelve months (the last block of five), the blocks reordered at random and the sequence truncated to 65; 10{,}000 draws with a fixed seed. At most $65 \times 6! = 46{,}800$ distinct label sequences exist, so the $p$-values are resolved to about $10^{-4}$ while the number of distinct episodes they rearrange is the small number above; the test addresses overlap, not the small count of episodes. The 53-month paired fundamental-minus-attention series uses the same procedure with five blocks (the last of five months), at most $53 \times 5! = 6{,}360$ distinct sequences (10{,}000 draws therefore repeat sequences), and terciles re-cut on those months at trailing return $-10.7\%$ and $+11.8\%$ and realized volatility 23.8\% and 29.6\%.
\end{minipage}
\end{table}

\begin{table}[!htbp]
\centering
\caption{Every channel on the same state grid, twelve months, 65 months (Section~\ref{sec:state}): mean neutral IC or neutral gain in the low, middle and high tercile of each state at formation, the high-minus-low difference and its two-sided $p$ from 10{,}000 circular block permutations (twelve-month blocks). Gains are paired against the market-only ranking; polarity and text rows are the series' own neutral IC. The fundamental-minus-attention row is the paired difference of the two gain series on the 53 months from 2021 (attention is first admitted in the 2021 fold), with terciles cut on those months, so it is not the arithmetic difference of the two rows above it; the Section~\ref{sec:state} contrasts are this row. Reading: fundamentals are $+0.046$ higher in high-volatility months ($p$ 0.003) and $+0.034$ after rallies ($p$ 0.049); the text head $+0.054$ after rallies ($p$ 0.014) and its identity placebo $+0.077$ in high-volatility months ($p$ 0.006); attention $-0.031$ after rallies ($p$ 0.008) and $-0.025$ in high volatility ($p$ 0.031); the clean social score $-0.054$ in high-volatility months ($p$ 0.016); the market score's own neutral IC does not vary by state ($p \geq 0.12$).}
\label{tab:stategrid}
{\scriptsize\setlength{\tabcolsep}{3pt}
\resizebox{\linewidth}{!}{\begin{tabular}{@{}lrrrrrrrrrrr@{}}
\toprule
 & & \multicolumn{5}{c}{Trailing 12m XBI return} & \multicolumn{5}{c}{Realized 12m volatility} \\
Series & Mean & Low & Mid & High & High$-$Low & $p$ & Low & Mid & High & High$-$Low & $p$ \\
\midrule
Market-only IC & 0.063 & 0.070 & 0.078 & 0.041 & $-0.029$ & 0.13 & 0.073 & 0.052 & 0.063 & $-0.010$ & 0.64 \\
Full hybrid gain & 0.030 & 0.029 & 0.021 & 0.040 & $+0.011$ & 0.61 & 0.014 & 0.029 & 0.047 & $+0.033$ & 0.07 \\
Fundamental gain & 0.022 & 0.009 & 0.015 & 0.043 & $+0.034$ & 0.05 & 0.002 & 0.018 & 0.048 & $+0.046$ & 0.003 \\
Attention gain & 0.021 & 0.039 & 0.016 & 0.009 & $-0.031$ & 0.008 & 0.032 & 0.026 & 0.007 & $-0.025$ & 0.03 \\
Fundamental minus attention, paired, 53 months from 2021, terciles re-cut & $-0.009$ & $-0.036$ & $-0.007$ & 0.014 & $+0.049$ & 0.023 & $-0.033$ & $-0.018$ & 0.021 & $+0.054$ & 0.011 \\
Scientific (structured) gain & $-0.001$ & $-0.000$ & $-0.000$ & $-0.001$ & $-0.000$ & 0.68 & $-0.000$ & $-0.000$ & $-0.001$ & $-0.001$ & 0.43 \\
Frontier polarity, news (Claude rating) & 0.006 & 0.021 & 0.001 & $-0.005$ & $-0.026$ & 0.34 & 0.007 & 0.017 & $-0.007$ & $-0.014$ & 0.58 \\
Frontier chain, social (FinBERT) & $-0.004$ & $-0.013$ & 0.011 & $-0.009$ & $+0.003$ & 0.92 & 0.003 & $-0.012$ & $-0.004$ & $-0.007$ & 0.78 \\
Clean polarity, news (GPT-2) & 0.025 & 0.040 & 0.000 & 0.032 & $-0.009$ & 0.84 & 0.026 & 0.022 & 0.025 & $-0.000$ & 0.99 \\
Clean polarity, social (GPT-2) & $-0.006$ & 0.007 & 0.003 & $-0.028$ & $-0.034$ & 0.17 & 0.021 & $-0.007$ & $-0.033$ & $-0.054$ & 0.016 \\
Clean polarity, summaries (GPT-2) & $-0.007$ & $-0.017$ & 0.015 & $-0.019$ & $-0.003$ & 0.95 & $-0.005$ & 0.007 & $-0.023$ & $-0.019$ & 0.60 \\
Text tower (2019 model, news-and-social panel) & 0.029 & $-0.002$ & 0.037 & 0.052 & $+0.054$ & 0.014 & 0.013 & 0.031 & 0.043 & $+0.030$ & 0.22 \\
Its identity placebo (same panel) & 0.042 & 0.018 & 0.043 & 0.065 & $+0.047$ & 0.10 & 0.009 & 0.031 & 0.086 & $+0.077$ & 0.006 \\
\bottomrule
\end{tabular}}}
\par\medskip\begin{minipage}{\linewidth}\footnotesize\raggedright\setlength{\parskip}{1pt}
\par The high-yield-spread state and the one-month grid are omitted here; on that state the full gain is $+0.039$ higher in wide-spread months ($p$ 0.04) and no channel's composition moves. The text-tower rows are from the first text panel (news and social text only, GPT-2 representation, 45{,}670 company-months), on which the state analysis was run; the all-source panel of Table~\ref{tab:texttower} was not re-run on the grid.
\end{minipage}
\end{table}

\FloatBarrier
\section{Lenses on Text Content}
\label{app:lenses}
Table~\ref{tab:lenses} reports the six lenses of Section~\ref{sec:clean}; the constructions and the decision rule are stated there, and the pre-2020 model's companion values are in the table's note.

\begin{table}[!htbp]
\centering
\caption{Six lenses on text content, Qwen3-Embedding representation, twelve months unless stated (three months in parentheses where it differs in kind): the lens, what a positive result would mean, the lens's own neutral IC with $t$, the identity placebo's, and the change in neutral IC from adding the lens's ranking to the hybrid (leave-one-year-out weight; $t$). L4 uses the clean rater's per-record scores as screened columns rather than a head. L6c is the event-month split of L6b.}
\label{tab:lenses}
{\scriptsize\setlength{\tabcolsep}{3pt}
\begin{tabular}{>{\raggedright\arraybackslash}p{0.25\linewidth}>{\raggedright\arraybackslash}p{0.12\linewidth}>{\raggedright\arraybackslash}p{0.11\linewidth}>{\raggedright\arraybackslash}p{0.07\linewidth}>{\raggedright\arraybackslash}p{0.17\linewidth}>{\raggedright\arraybackslash}p{0.15\linewidth}}
\toprule
Lens & Would mean & Neutral IC ($t$) & Identity placebo & Added to hybrid & Verdict \\
\midrule
L1 Change: vector minus the company's past mean & month's news predicts & $-0.004$ ($-0.8$) & $-0.002$ & 0 (weight 0) & no content \\
L2 Residual: head on return minus market-score fit & text holds what numbers miss & $+0.024$ (2.8) & $+0.025$ & $+0.002$ (1.1) & persistent signal again \\
L3 Magnitude: head on $|$excess return$|$ & text predicts event risk & $0.144$ (8.2)$^{\dagger}$ & $0.182$ & -- & persistent signal; trailing volatility 0.329 \\
L4 Tail: negative-record share, 90 days (screened columns) & rare bad news predicts & $+0.002$ to $+0.013$ ($\leq1.3$) & -- & not admitted & no content; counts flip sign \\
L5 Gaps: rows with no non-market coverage & text fills gaps & $+0.081$ (market 0.085) & -- & $+0.007$ (0.4); 3m $-0.042$ ($-1.9$) & no \\
L6a Attention: text IC by coverage tercile & content needs coverage & $0.025 / 0.032 / 0.046$ & -- & -- & rises with coverage, as a persistent signal would \\
L6b Source: scientific text alone, 3m & content lives in one source & $+0.040$ (5.4) & $+0.035$ & $+0.010$ (2.8; $[+0.006, +0.015]$); raw $+0.003$ (2.2); placebo $-0.001$ & the one lead (12m: $+0.003$, 1.2) \\
L6b Source: news alone (6\% coverage); social alone & & $-0.003$; $-0.001$ (3m) & $+0.042$; $+0.009$ & $-0.007$; $-0.002$ & nothing \\
L6c Event months: scientific text, 3m, rows whose window holds a trial, FDA or advisory record (3{,}170 rows, 16\%, 65 months) & content sits where events are & $+0.042$ (2.9) & $+0.009$ & $+0.064$ (2.9; $[+0.024, +0.103]$); placebo $+0.024$ (1.0); raw $+0.011$ (0.4) & hybrid alone $-0.023$ on these rows; $w=1$ in every fold \\
L6c Other months, 3m (16{,}543 rows, 72 months); event months, 12m (2{,}688 rows, 58 months) & & $+0.030$ (4.1); $+0.008$ (0.6) & $+0.032$; $+0.046$ & $+0.005$ (1.3; $[+0.000, +0.010]$); $+0.003$ (0.1; $[-0.039, +0.041]$) & nothing beyond the all-rows lead \\
\bottomrule
\end{tabular}}
\par\medskip\begin{minipage}{\linewidth}\footnotesize\raggedright\setlength{\parskip}{1pt}
\par 2019-model (GPT-2) companion values, same lenses and placebo: L1 change $-0.006$; L2 residual $+0.025$ against a placebo of $+0.013$, adding $+0.002$; L3 magnitude 0.149 against 0.173; L5 gap rows $+0.029$ ($t$ 1.5); L6a text IC 0.043 / 0.042 / 0.040 across attention terciles, flat where the 2025 model's rose; L6b scientific text alone $+0.030$ at three months, adding $+0.004$ ($t$ 2.0) to the hybrid. Verdicts identical. $^{\dagger}$Spearman with the realized absolute excess return. L1 uses the expanding mean of a company's prior months' vectors (at least three), 21{,}422 twelve-month rows; its within-month placebo is $+0.000$. L2 is walk-forward (the residual is after a within-month rank regression on the market score, which exists from 2015); a residual on the hybrid's own prediction can only be fitted leave-one-year-out across the evaluation years and then carries company-level future information, so it is not reported as a lens (its head and its placebo are equal at $+0.058$, and it adds $+0.003$, $t$ 0.7, to the hybrid). L5 gap rows are 7\% of twelve-month and 35\% of three-month rows with text. Eighteen lens-by-horizon tests; the two-sided Bonferroni value is 2.99.
\end{minipage}
\end{table}

\FloatBarrier
\section{Pre-Specified Hypotheses and Outcomes}
\label{app:prespec}

Each hypothesis (Table~\ref{tab:prespec}) was written into the manuscript repository before the computation it names was run; the commit history is the record (no external registry), and Table~\ref{tab:chrono} gives it, with the gap between specification and result where it is short. Results contradicting a hypothesis are reported as such. Two facts about the chronology bear on how the hypotheses should be read. The state pattern that H5 formalizes was first seen in an earlier, unpublished year-by-year analysis of the same system (September 2026), so H5 is a confirmation of an observed pattern on a new comparator, not a prediction. H8 and the six lenses were specified minutes before their scripts ran; the specification is pre-registered in the weak sense that the decision rule was fixed before the result was seen, and the analyses are exploratory in every other sense.

\paragraph{Search inventory and multiplicity.} The search disclosed in this paper comprises: sixteen market architectures on the 2015--2018 contest; the screen's gates; 184 firm-level columns plus three macro variables, across three horizons: 561 tests per fold (Bonferroni 3.92, local to that family); twelve GPT-2-rated columns and nine fact columns screened in their own channel pools, not added to the full model's; two text encoders; two shuffles; average and learned blends on two base rankings; six lenses with three source arms (eighteen tests, Bonferroni 2.99, local to that family); event-month and attention-tercile subgroups; four state definitions, two of them pre-specified; and two- and three-way regime blends over a 66-triple grid. Each multiplicity correction is local to the family it names; no correction spans the full inventory, and the headline ledger $t$-statistics, the state permutation $p$-values and the scientific-text increment are reported unadjusted for it. Blend weights chosen leave-one-year-out are retrospective, fitted on the other evaluation years. The all-channel state grid for the two pre-specified states is Table~\ref{tab:stategrid}; the high-yield-spread state and the one-month grid are available with the code.

\begin{table}[!htbp]
\centering
\caption{Chronology of specification and result, from the repository's commit timestamps (local time).}
\label{tab:chrono}
{\scriptsize
\begin{tabular}{@{}p{2.9cm}p{2.9cm}p{3.3cm}p{4.6cm}@{}}
\toprule
Analysis & Specified & Result committed & Note \\
\midrule
H1--H3 (clean chain, cutoff) & 1 Oct 2026, 23:56 & H2 frontier rows 2 Oct 19:08; clean-chain rows 3 Oct 06:01 & Rater trained and scored 1--3 Oct between the two. \\
H4 (ledger bound), H5 (regime blend), H6 (text tower) & 2 Oct, 00:09--02:11 (plan v2, v3) & H4, H5: 2 Oct 15:51; H6: 3 Oct 01:18 on the news-and-social panel (GPT-2 embeddings 2--3 Oct); all-source panel 4 Oct 15:53 (2025 model) and 5 Oct 00:05 (2019 model) & H5's state pattern was observed in an unpublished year-by-year analysis in September 2026. \\
Within-company placebo & 3 Oct, 00:20, after a partial-coverage dry run of the text tower & 3 Oct 01:18 & Added before the full-coverage run; the dry run had shown skill without the placebo. \\
H7 (prompted forecasting) & 3 Oct, 19:05 (experiment built, not run) & 4 Oct 14:03 & Two pilots (3 Oct 19:46) changed the token budget and reasoning effort only; the prompt text was fixed at specification. \\
H8 (three-way regime blend) & 4 Oct, 16:08 & 4 Oct 16:12 & Script committed 16:09; four minutes between specification and result. \\
Six lenses and decision rule & 4 Oct, 16:56 & 4 Oct 17:13; source split 18:54--18:57 & Scientific-text lead found in the source-split arm. \\
Event-month split (L6c) & Not pre-specified & 4 Oct 22:20 (ad hoc); saved script 5 Oct & Post hoc subgroup comparison. \\
2019-model (GPT-2) rows of the text tower and lenses & same designs & 5 Oct 00:05 & Same scripts, second representation. \\
\bottomrule
\end{tabular}}
\end{table}

\begin{table}[!htbp]
\centering
\caption{Pre-specified hypotheses, specification dates and outcomes.}
\label{tab:prespec}
{\footnotesize
\begin{tabular}{>{\raggedright\arraybackslash}p{0.04\linewidth}>{\raggedright\arraybackslash}p{0.46\linewidth}>{\raggedright\arraybackslash}p{0.10\linewidth}>{\raggedright\arraybackslash}p{0.32\linewidth}}
\toprule
 & Hypothesis & Specified & Outcome \\
\midrule
H1 & No clean-chain (GPT-2) column is admitted by the $|t|\geq4$ screen in any fold--horizon pair. & 2026-10-01 & Met: 0 of the 12 GPT-2-rated columns in 18 pairs; closest $|t|$ 3.39 (summary rating, 12m, 2020 fold). \\
H2 & The frontier-minus-clean IC gap is larger before the frontier rater's cutoff than after it. & 2026-10-01 & Not supported at 1m and 3m on any pair (Claude vs FinBERT $+0.007$ $[-0.022, +0.035]$ and $-0.022$; vs GPT-2 $-0.005$, $-0.018$; GPT-5.4-mini vs FinBERT $-0.013$, $-0.036$; vs GPT-2 $-0.024$, $-0.040$). At 12m the GPT-2 pairs fall by $+0.022$ and $+0.031$ but the clean--clean placebo falls by $+0.070$ ($t$ 2.8): not identified. \\
H3 & On post-cutoff months, the clean rater on summaries does not beat the clean rater on raw records. & 2026-10-01 & Met at 1m and 3m (summary minus raw $-0.020$ and $-0.032$ IC after the Claude cutoff; $-0.012$, $-0.030$ after the GPT-5.4-mini cutoff); at 12m the summary rating leads by $+0.130$ on fifteen overlapping windows in which the raw-record IC is $-0.089$, the same unstable cells as H2. \\
H4 & The clean-chain ledger row is indistinguishable from zero; the fundamental row is not. & 2026-10-02 & Met: fundamental $+0.022$ (12m, $t$ 2.8); GPT-2-rated row zero by construction in every pair, post hoc magnitude bounds 0.044 / 0.081 / 0.063. \\
H5 & The LOYO regime blend's 12m neutral IC exceeds the static blend's. & 2026-10-02 & Met: $+0.013$ ($t$ 2.4 and 2.2 on the two states); matches the full model ($+0.004$, $t$ 0.8). \\
H8 & The three-way regime blend (market-only, hybrid, LLM text ranking; weights by state tercile, LOYO) has 12m neutral IC above the hybrid's. & 2026-10-04 & Not met under the paper's two-sided 5\% Newey--West convention; the recorded rule names only the direction (one cell, the high-yield scientific-text variant, has a positive bootstrap interval $[+0.001, +0.024]$ with $t$ 1.7 and raw $-0.031$): all-text ranking $-0.004$ to $+0.002$ against the hybrid on the single states ($t$ $-1.2$ to $+0.4$) and $-0.009$ on the joint state ($t$ $-2.1$); scientific-text ranking $+0.002$ to $+0.011$ on the single states ($t\leq1.7$) and $-0.003$ on the joint state. At 3m the scientific-text blend on text rows adds $+0.006$ to $+0.009$ ($t\leq2.5$), reported as a lead. 2019 ranking: $-0.007$ to $-0.001$ ($t\leq1.9$), not met. \\
H6 & The text-only ranking (frozen representation, walk-forward head) has neutral IC indistinguishable from zero and adds nothing to the numeric model. & 2026-10-02 & First half not met at 3m and 12m for both generations ($+0.030$ / $+0.039$ for the 2019 model, $+0.019$ / $+0.034$ for the 2025 model; zero at 1m); the within-company shuffle (added 2026-10-03 after a partial-coverage dry run) matches it, exceeding it in five of six cells (the exception: the 2019 model at 12m, $+0.039$ against $+0.036$), so the skill is persistent company-level language; second half not met at 3m ($+0.004$, $[+0.002, +0.005]$; scientific text $+0.010$) and not rejected at 12m, where adding either ranking changes the hybrid's IC by $+0.003$ and a plain average lowers it. \\
H7 & Prompted forecasting (specified before the run): the revealed-minus-blind IC gap is larger before the model's cutoff than after it; the blind score adds nothing to the hybrid. & 2026-10-03 & First half directionally met: the gap declines at every horizon ($+0.017$, $t$ 1.7 at 12m; $+0.012$, $t$ 2.5 at 3m; $+0.008$, $t$ 0.8 at 1m), nominally significant at three months only. Second half met: blend weight 0 at twelve and one months, 0 or 0.1 at three. \\
\bottomrule
\end{tabular}}
\end{table}

\end{document}